\documentclass[journal]{IEEEtran}

\makeatletter
\def\ps@headings{%
\def\@oddhead{\mbox{}\scriptsize\rightmark \hfil \thepage}%
\def\@evenhead{\scriptsize\thepage \hfil \leftmark\mbox{}}%
\def\@oddfoot{}%
\def\@evenfoot{}}
\makeatother 

\IEEEoverridecommandlockouts

\usepackage{caption}
\usepackage{algorithm}
\usepackage{clrscode}
\usepackage{bbm}
\usepackage{amsfonts}
\usepackage[dvips]{graphicx}
\usepackage{times}
\usepackage{cite}
\usepackage{amsmath}
\usepackage{array}
\usepackage{amssymb}

\usepackage{stfloats}
\usepackage{slashbox}
\usepackage{graphicx}
\usepackage{subfigure}
\usepackage{footnote}
\usepackage{amsthm}
\usepackage{booktabs}
\usepackage{array}
\usepackage{algorithmic}
\usepackage{algorithm}
\usepackage{subeqnarray}
\usepackage{cases}
\usepackage{threeparttable}
\usepackage{color}
\usepackage{multirow}
\usepackage{xcolor}
\usepackage{epstopdf}
\usepackage{bm}
\usepackage{dsfont}
\usepackage{enumerate}
\usepackage{float}
\usepackage{hyperref}
\newcommand{\tabincell}[2]{\begin{tabular}{@{}#1@{}}#2\end{tabular}}
\begin{document}

\title{Low Complexity Iterative Receiver Design \\ for Sparse Code Multiple Access}

\author{Fan~Wei and
        Wen~Chen,~\IEEEmembership{Senior Member,~IEEE}% <-this % stops a space
\thanks{This paper was presented in part at the IEEE Global Communications Conference (GLOBECOM), Washington DC, USA, December 2016.}
\thanks{This work is supported by NSFC \#61671294, National 863 Project \#2015AA01A710, STCSM Key Fundamental Project \#16JC1402900, and GXNSF \#2015GXNSFDA139037.}
\thanks{Fan Wei and Wen Chen are with Shanghai Key Lab of Navigation and Location Based Services,
Shanghai Jiao Tong University, China, and School of Electronic Engineering and Automation, Guilin University of Electronic Technology, China. (e-mail:  \{weifan$89$; wenchen\}@sjtu.edu.cn).}% <-this % stops a space
}

\markboth{IEEE Transactions on Communications}%
{Submitted paper}

\maketitle

\begin{abstract}
%\boldmath
Sparse code multiple access (SCMA) is one of the most promising methods among all the non-orthogonal multiple access techniques in the future 5G communication. Compared with some other non-orthogonal multiple access techniques such as low density signature (LDS), SCMA can achieve better performance due to the shaping gain of the SCMA codewords. However, despite of the sparsity of the codewords, the decoding complexity of the current message passing algorithm (MPA) utilized by SCMA is still prohibitively high. In this paper, by exploring the lattice structure of SCMA codewords, we propose a low complexity decoding algorithm based on list sphere decoding (LSD). The LSD avoids the exhaustive search for all possible hypotheses and only considers signal within a hypersphere. As LSD can be viewed a depth-first tree search algorithm, we further propose several methods to prune the redundancy visited nodes in order to reduce the size of the search tree. Simulation results show that the proposed algorithm can reduce the decoding complexity substantially while the performance loss compared with the existing algorithm is negligible.
\end{abstract}


\IEEEpeerreviewmaketitle

\section{Introduction}
\IEEEPARstart{T}{he} growing demand for wireless data transmission has inspired the researchers' interests on the next generation wireless communication network. Non-orthogonal multiple access technique along with massive multiple-input multiple-output (MIMO), ultra-dense radio networking, all-spectrum access, device to device communications and so on are among some key technologies in 5G wireless networks~\cite{01}. Compared with orthogonal multiple access techniques such as code division multiple access (CDMA) or orthogonal frequency division multiple access (OFDMA) used in the current networks, non-orthogonal multiple access, due to its high efficient utilization of resource, is more attractive in 5G scenarios, such as the internet of things, that has massive connectivity requirements.

The investigation of non-orthogonal multiple access technology has been done by a number of researchers. Motivated by the design of CDMA chip sequences, Hoshyar \emph{et al}.~\cite{02} proposed a novel structure which is called low density signature (LDS). Instead of the conventional approach for designing orthogonal or near orthogonal spreading signatures to avoid interference, multiuser interference is allowed in the LDS. The difference between LDS and CDMA is that chip sequences in LDS are sparse so that for each user only small number of chips are used to spread data. The sparsity of spreading signature can reduce the number of interfering users in each chip so that the near optimal message passing algorithm (MPA)~\cite{03,04} can be used to decode data for each user. Later, Beek \emph{et al}.~\cite{05} discussed the design of LDS spreading signature. It was shown that by optimizing the rotation angle, unique decodability for each user can be guaranteed and the distance spectrum of the system can also be optimized. LDS combined with orthogonal frequency division multiplexing (OFDM), termed as LDS-OFDM, was firstly introduced in~\cite{06}. Due to its capability to explore the frequency domain diversity, LDS-OFDM was proved to have a superior performance than OFDMA. The capacity region of LDS on multiple access channel (MAC) was analysed in~\cite{07} where the weighted sum rate MAC capacity and LDS MAC capacity were compared. In general, LDS capacity is lower than MAC capacity due to its suboptimal multiuser detection.

Sparse code multiple access (SCMA) is another non-orthogonal multiple access technology introduced by Nikopour \emph{et al}.~\cite{08,09}. Instead of the simple repetition of quadrature amplitude modulation (QAM) symbols in LDS, coded bits are directly mapped to multi-dimensional complex lattice point (called a codeword) in SCMA~\cite{a2}. The data in each dimension of the codeword is modulated to an orthogonal resource, e.g., an OFDMA subcarrier and then transmitted to the radio channels. Similar to LDS, the codewords in SCMA are sparse since only a small portion of dimensions are used to transmit data. Also, the system is overloaded as in LDS due to the number of multiplexed codewords are more than the number of OFDMA subcarriers. Therefore, both schemes can use a factor graph to represent their structures. However, the difference to LDS is that the codewords in SCMA are essentially multi-dimensional complex lattice points, thus shaping gain of multi-dimensional complex lattice points makes SCMA achieve a superior performance compared with LDS.

\begin{figure*}[t]
  \centering
  \includegraphics[width=7.1in,height=2.0in]{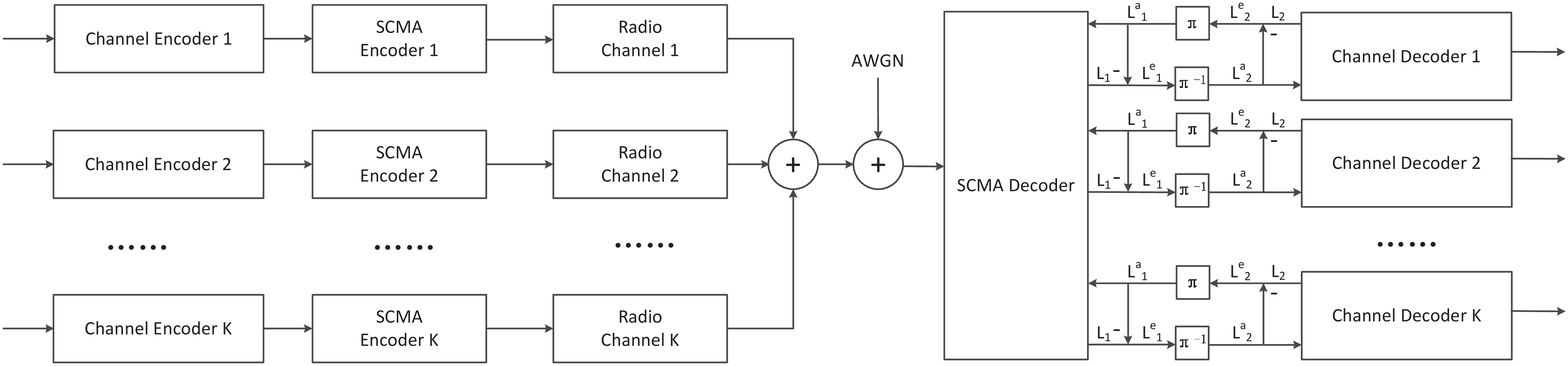}
  \caption{Block diagram of uplink SCMA system}\label{Fig.1}
\end{figure*}

In general, SCMA has two distinctive characteristics compared with other multiple access techniques. The overloaded feature of SCMA makes non-orthogonal multiple access (and correspondingly massive connectivity) possible and the sparsity of the codewords renders the use of the suboptimal message passing algorithm. However, despite the sparsity of the codewords in SCMA, the detection of received signals are still time consuming. The brute force search for all possible signals on each OFDMA subcarrier brings the exponential computational complexity in MPA.

To reduce the decoding complexity, there has been a number of improvements on the conventional MPA. In the design of SCMA codebook~\cite{09}, rotation of the QAM symbols is introduced to improve the performance in fading channels. While there exists some optimal rotation angle to get an optimized product distance~\cite{09a}, certain suboptimal rotation angles could result in a low number of projections (LNP)~\cite{09b} in each OFDMA tone.~\cite{10} has discussed MPA in log domain and pointed out that the Log-MPA could convert considerable number of multiplications to summations, thus saves the detection time effectively. Based on partial marginalization (PM), a modified MPA is proposed in~\cite{11}, where during the iteration, the codewords from part of users are chosen as reference symbols and not updated in the remaining iterations. Therefore, by tailoring the edges in the Forney factor graph, the collision users on each subcarrier can be reduced. Further, weighted message passing is introduced in~\cite{13} to reduce the number of MPA iterations. In the weight setting, constellation points that are closer to the received signal will be distributed to a lager weight which ensues that the points with high probability will be even higher.

The above methods have offered us with several different perspectives to reduce the decoding complexity of SCMA. In this paper, motivated by the fact that SCMA codewords are essentially multi-dimensional complex lattice points, we propose a low complexity algorithm based on sphere decoding~\cite{14,15,16}. The sphere decoding is an efficient method to reduce the complexity of \emph{maximum likelihood} (ML) detection, which avoids the brute force search for all possible signals but limits the search space within a given hypersphere. Sphere decoding has been shown to be efficient in dealing with lattice codes~\cite{17,18,19}, MIMO detection~\cite{16,20,21,22,23}, network coding~\cite{24,25} and so on. For the decoding of SCMA codewords, we will show that the soft information updated during the MPA iteration can be effectively approximated through using the candidate list searched by the list sphere decoder~\cite{16,a1}. In addition, as the sphere decoding can be viewed as a tree search algorithm, some node pruning techniques are further developed to reduce the redundancy visited nodes during the tree search process.

The remainder of this paper is organized as follows: in Section II an uplink SCMA system model is introduced. The lattice structure of SCMA codewords is formulated in Section III, where we also briefly introduce the MPA and iterative decoding and detection (IDD) of SCMA. Section IV discusses the implementation of list sphere decoding (LSD) based MPA as well as some node pruning techniques. Simulation results are presented in Section V and the final conclusion is summarised in Section VI.

Throughout this paper, the following notations will be used. Lowercase letters $x$, bold lowercase letters $\mathbf{x}$ and bold uppercase letters $\mathbf{X}$ denote scalars, column vectors and matrices, respectively. The superscripts $(\cdot)^{*}$ denotes the complex conjugate and $(\cdot)^{T}$ denotes matrix transpose, while $(\cdot)^{\dag}$ denotes conjugate matrix transpose. $\mbox{diag}(\mathbf{x})$ is the diagonal matrix with the diagonal entries being vector $\mathbf{x}$. $\xi \setminus k$ means the set $\xi$ with element $k$ being excluded. $C(n,k)$ is the number of $k$-combinations from a given set that has $n$ elements. We define $\mbox{sign}(x) = 1$ if $x>0$ and $\mbox{sign}(x) = -1$ otherwise.

\section{SCMA System Model}
The block diagram of synchronous uplink SCMA system is depicted in Fig.~\ref{Fig.1}, In the uplink, an SCMA encoder corresponds to one layer of the SCMA system, and each user may occupy one or more SCMA layers. In this paper, we assume that each user has only one SCMA layer. The transmitted signal from each layer is multiplexed to some orthogonal resources, e.g. OFDMA subcarriers. Due to the non-orthogonal feature of SCMA system, the total number of layers may be more than the number of orthogonal resources. Thus, SCMA is an overloaded system. The superimposed signal from each layer is processed by an SCMA decoder, which uses the near optimal message passing algorithm. The log-likelihood ratios (LLR) of coded bits are computed there and sent to the channel decoders, from which the final decoded bits are obtained.

\subsection{SCMA Transceiver Structure}
For each user $k$, the SCMA encoder can be defined as a mapping of coded bits $\mathbf{b_{k}}$ to a multi-dimensional complex codeword $\mathbf{x_{k}}$, namely $\mathbf{x_{k}} = f(\mathbf{b_{k}})$, $f: \mathbb{B}^{\log_{2}(M)}\rightarrow \mathcal{X},\ \mathcal{X}\in \mathbb{C}^{N}$, where $N$ is the dimension of the complex codeword and $M$ is the cardinality of the codebook, i.e. $|\mathcal{X}| = M$. The data in each dimension are then modulated to an OFDMA subcarrier and transmitted to the air interface. In SCMA, the multi-dimensional codewords are sparse, in other words, only $P < N$ dimensions are used to transmit data while the other $N-P$ dimensions are set to zeros. The sparsity of the codewords can reduce the interference from other users and makes the near optimal MPA detection possible. Due to the sparsity of the codewords, we can rewrite the SCMA encoder as $f : \equiv \mathbf{V}_{k}g$, where encoder $g:\mathbb{B}^{\log_{2}(M)}\rightarrow \mathcal{C}, \ \mathcal{C}\in \mathbb{C}^{P}$ encodes the binary input data of each user to a $P$-dimensional nonzero codeword while matrix $\mathbf{V}_{k}\in \mathbb{B}^{N\times P}$ transforms the $P$-dimensional nonzero codeword to an $N$-dimensional sparse codeword.

The received signal after synchronous layer multiplexing at base station can be expressed as
\begin{equation}\label{01}
  \mathbf{y} = \sum_{k=1}^{K} \mbox{diag}(\mathbf{h}_{k}) \mathbf{x}_{k} + \mathbf{z},
\end{equation}
where vectors $\mathbf{x}_{k}$, $\mathbf{h}_{k}$ are the SCMA codeword and fading channel of user $k$, $\mathbf{z}$ is the additive complex Gaussian noise vector with distribution $\mathcal{CN}(\mathbf{0}, \sigma^{2}\mathbf{I})$. As $K$ users are multiplexed to $N$ orthogonal resources, the overloading factor of the system is $\frac{K}{N}$.

\subsection{Factor Graph Representation}
Since the SCMA codewords are sparse, only a few users collide over the OFDMA subcarrier $n$. In order to capture this feature of SCMA, we introduce an indicator vector $\mathbf{f}_{k}\in \mathbb{B}^{N}$ for each user $k$, where the $n$th elements $f_{n,k}$ is defined as
\begin{equation}\label{02}
   f_{n,k} = \left\{
              \begin{array}{ll}
                0, & x_{n,k}= \hbox{0;} \\
                1, & x_{n,k}\neq \hbox{0.}
              \end{array}
            \right.
\end{equation}
for $k = 1,2,...,K$, and the corresponding indicator matrix is given by $\mathbf{F} = [\mathbf{f}_{1},\mathbf{f}_{2},...,\mathbf{f}_{K}]$. In matrix $\mathbf{F}$, the set of nonzero entries in each row correspond to the users who collide over the same subcarrier while the ones in each column denote the set of subcarriers that user $k$ transmits its data. Due to the sparsity of SCMA codewords, matrix $\mathbf{F}$ is also sparse. For each user $k$, the subcarriers through which the data are transmitted can be determined by a mapping matrix $\mathbf{V}_{k}$. The design of mapping matrix $\mathbf{V}_{k}$ has been discussed in~\cite{08,09}. It can be obtained by inserting $N-P$ all-zero rows into an identity matrix $\mathbf{I}_{P}$. For instance, when $P=2$, $N=4$, the mapping matrix of user $k$ may be
\begin{equation*}
  \mathbf{V}_{k}=\begin{bmatrix}
  1 & 0 \\
  0 & 0 \\
  0 & 1 \\
  0 & 0 \\
\end{bmatrix}.
\end{equation*}
Clearly, we can have at most $K=C(N,P)$ different matrices for a given $N$ and $P$. With matrix $\mathbf{V}_{k}$, the indicator vector $\mathbf{f}_{k} $ of user $k$ can be determined by $\mathbf{f}_{k} = \mbox{diag}(\mathbf{V}_{k}\mathbf{V}_{k}^{T}) $.

\begin{figure}
  \centering
  \includegraphics[width=3.4in,height=1.75in]{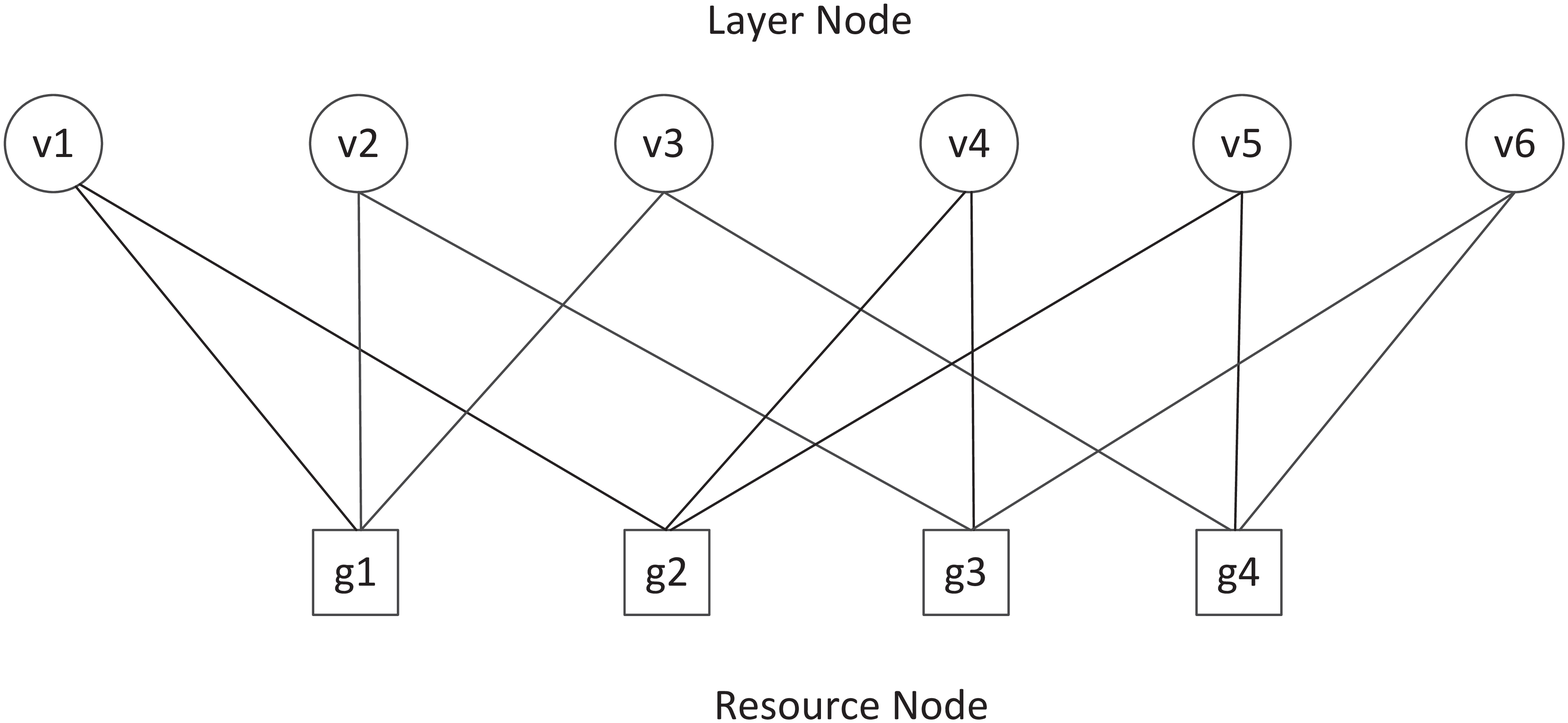}
  \caption{Factor graph of SCMA with $K=6$, $N=4$ and $P=2$ }\label{Fig.2}
\end{figure}

The structure of indicator matrix $\mathbf{F}$ can be represented by a Forney factor graph as the case in low density parity check (LDPC) code. In the graph, layer node $k$ and resource node $n$ are connected if and only if $f_{n,k} = 1$. We define two sets $\xi_{n} = \{k|f_{k,n} \neq 0\}$ for $n = 1, ..., N$ and $\zeta_{k} = \{n|f_{k,n} \neq 0\}$ for $k = 1, ..., K$ which correspond to the set of users collide over subcarrier $n$ and the set of subcarrers occupied by user $k$ respectively. In each subcarrier, there are $d_{c}$ collision users, i.e. $|\xi_{n}| = d_{c}$, which is given by $d_{c} = \frac{KP}{N}$. The overloading factor of SCMA system is $\lambda = \frac{K}{N}$. Fig.~\ref{Fig.2} shows a factor graph with $N=4, P=2, K=6, d_{c}=3$ and $\lambda = 150\%$, and~\eqref{02a} is the corresponding indicator matrix.
\begin{equation}\label{02a}
  \mathbf{F} = \left[
                   \begin{array}{cccccc}
                     1 & 1 & 1 & 0 & 0 & 0 \\
                     1 & 0 & 0 & 1 & 1 & 0 \\
                     0 & 1 & 0 & 1 & 0 & 1 \\
                     0 & 0 & 1 & 0 & 1 & 1 \\
                   \end{array}
                 \right].
\end{equation}

\section{Lattice Structure of SCMA Codewords \\ and Iterative Multiuser Detection}
In this section, we introduce the design of SCMA codebook and clarify the lattice structure of SCMA codewords. This lattice structure is shown to facilitate the implementation of low complexity sphere decoding discussed in section IV. Before elaborating the new decoding algorithm, MPA and iterative multiuser detection for SCMA will be introduced firstly in this section.

\subsection{SCMA Codebook Design}
The design of SCMA codebook has been discussed in literatures~\cite{08,09}. In general, SCMA codewords can be constructed from any complex constellation points with minimum energy $E_{s}$ for a given Euclidean distance $d_{min}$, i.e. constellations with minimized figure of merit $\eta = \frac{E_{s}}{d_{min}}$. For codewords with higher rate,~\cite{09} has proposed a structured multi-step suboptimal approach to construct the complex constellation. This multi-step suboptimal approach can be formulated as below
\begin{equation}\label{03}
  \mathcal{V^{*},G^{*}} = \arg\max_{\mathcal{V,G}}\emph{m}(\mathcal{S}(\mathcal{V,G};K,M,P,N)),
\end{equation}
where \emph{m} is a given design criterion, $\mathcal{V}:=[\mathbf{V}_{k}]^{K}_{k=1}$ are the mapping matrices and $\mathcal{G}: = [\mathcal{G}_{k}]^{K}_{k=1}$ are the SCMA encoders that encode the incoming bits of each user to a $P$-dimensional nonzero codeword. The design of mapping matrix has been discussed in section II while in this section we will focus on the construction of $P$-dimensional complex constellation. Instead of designing $K$ different constellations for each layer, we could construct one mother constellation $\mathcal{G}$ and $K$ different constellation operators $[\pmb{\Delta}_{k}]^{K}_{k=1}$ so that the optimization problem can be formulated as
\begin{align}\label{04}
  \nonumber \mathcal{G}^{+},[\pmb{\Delta}^{+}_{k}]^{K}_{k=1} & = \arg\max_{\mathcal{G},\pmb{\Delta}_{k}} \\
   & \emph{m}(\mathcal{S}(\mathcal{V^{+}},\mathcal{G}=[(\pmb{\Delta}_{k})\mathcal{G}]^{K}_{k=1};K,M,P,N)).
\end{align}
As the optimal design criterion $\emph{m}$ is unknown, it is difficult to get a closed form optimization result for the above problem. Therefore, instead of the direct optimization, a suboptimal multi-stage approach is proposed in~\cite{08,09}, where a separate design of $P$-dimensional mother constellation $\mathcal{G}$ and $K$ different constellation operators $[\pmb{\Delta}_{k}]^{K}_{k=1}$ are implemented. From~\cite{08,09}, the $P$-dimensional mother constellation can be formed by Cartesian product of $P$ independent QAM symbols. A unitary rotation is further applied to the constellation. The goal of unitary rotation is to increase the modulation diversity of the constellation, i.e. the minimum number of distinct components between any two constellation points~\cite{09a}. In SCMA, the rotation can also induce dependency among the lattice dimensions and create power variation on different dimensions which lead to a better convergence of the MPA detector.

Let $\mathbf{u}_{2P}=(u_{1},...,u_{2P})$ be the equivalent $2P$-dimensional QAM constellation, where $u_{i} = \pm1,\pm3,...$, and $\mathbf{M}_{2P\times2P}$ be the unitary rotaion matrix. Then, the $P$-dimensional complex mother constellation point discussed above can be formulated as~\cite{09a}
\begin{equation}\label{05}
  \mathbf{x}_{P} = (\mathbf{E}_{r} + i  \mathbf{E}_{i}) \cdot \mathbf{M} \cdot \mathbf{u}_{2P},
\end{equation}
where matrices $\mathbf{E}_{r}$ and $\mathbf{E}_{i}$ are the $P\times2P$ matrices which select components from vector $\mathbf{u}_{2P}$ that corresponds to the real part and imaginary part of QAM symbols, respectively.

\begin{figure}
  \centering
  \includegraphics[width=3.4in,height=1.75in]{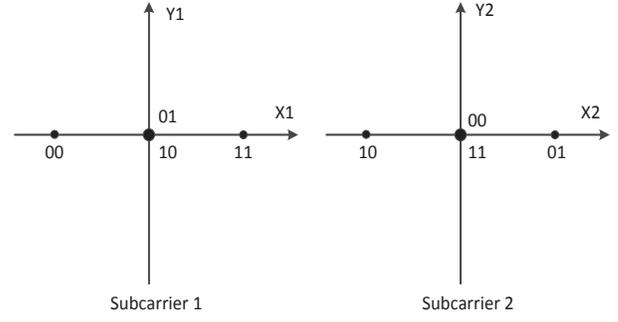}
  \caption{Low number of projections for 4 points SCMA}\label{Fig.2a}
\end{figure}

\begin{figure}
  \centering
  \includegraphics[width=3.4in,height=1.75in]{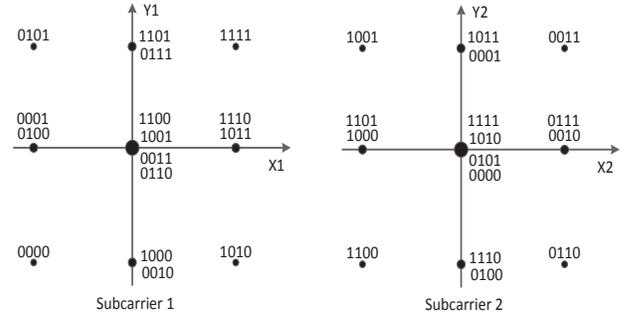}
  \caption{Low number of projections for 16 points SCMA}\label{Fig.2b}
\end{figure}

~\cite{09a} has discussed carefully to find the best rotation matrix $\mathbf{M}$. As a trade off between the performance and complexity, we can reduce the number of projections in each OFDMA tone by rotating the QAM symbols with some suboptimal rotation angles. For instance, for the two dimensional rotation matrix\footnote{In~\cite{09a}, the rotation constellation signal is written as $\mathbf{x}=\mathbf{u}\cdot\mathbf{M}$. Due to the formulation of~\eqref{05} in this paper, we make a transpose of the rotation matrix reported in~\cite{09a}.}
\begin{equation*}
  \mathbf{M}=
  \begin{pmatrix}
     a & -b \\
   b & a \\
  \end{pmatrix}.
\end{equation*}
Let $a=-b$ and $u_{i}\in$ binary phase shift keying (BPSK) signal. We can get the $4$ points constellation as shown in Fig.~\ref{Fig.2a}. Moreover, the constellation for $16$ points SCMA shown in Fig.~\ref{Fig.2b} can be obtained by using the rotation matrix
\begin{equation*}
 \mathbf{M} = \begin{bmatrix}
    a & -b & c & -d \\
    b & a & d & c \\
    -c & d & a & -b \\
    -d & -c & b & a \\
  \end{bmatrix},
\end{equation*}
with $a=c$ and $b=d=0$. The constellation shape is similar to that of Fig. $4$ in~\cite{09}. With the suboptimal rotation matrix $\mathbf{M}$, one can get a reduced number of projections in each non-zero dimension so that the decoding complexity can be reduced from $M^{d_{c}}$ to $m^{d_{c}}$ with $M>m$.

To construct different codebooks for different users, constellation operators are further needed. There are three types of operators, namely,
\begin{enumerate}
  \item Phase rotation: $\theta_{k} = (k-1)\frac{2\pi}{Md_{c}}, \forall k = 1,...,d_{c}$;
  \item Dimensional permutation;
  \item Complex comjugate.
\end{enumerate}
Detailed information about the construction of these operators can be found in~\cite{08,09}.

We will denote the above constellation operators as $\pmb{\Delta}_{k}$ for user $k$. To sum up, the non-zero $P$-dimensional SCMA codeword of user $k$ can be written as,
\begin{align}\label{06}
  \nonumber \mathbf{x}_{P,k} & = \pmb{\Delta}_{k}\cdot(\mathbf{E}_{r} + i \cdot \mathbf{E}_{i})\cdot \mathbf{M} \cdot
            \mathbf{u}_{2P,k} \\
                             & = \mathbf{G}_{k}\cdot \mathbf{u}_{2P,k},
\end{align}
where $\mathbf{G}_{k} =  \pmb{\Delta}_{k}\cdot(\mathbf{E}_{r} + i \cdot \mathbf{E}_{i})\cdot \mathbf{M}$. Clearly, each SCMA codeword is one point of the multi-dimensional lattice constellation. As we will see later, this lattice structure could be beneficial for utilizing some low complexity decoding algorithms.

\subsection{MPA and Iterative Detection and Decoding}
Iterative detection and decoding is a way to improve the bit error rate (BER) performance by exchanging the soft information between the detector and decoder iteratively. It has been used in MIMO detection extensively and was first proposed to SCMA in\cite{26}. In Fig.~\ref{Fig.1}, we illustrate the block diagram of IDD for SCMA.

For an SCMA decoder, the \emph{a posterior} \emph{L}-value of coded bit $b_{k}^{i}$ conditioned on the channel received value $y_{n}$ is given by,
\begin{equation}\label{06a}
  L_{1}(b_{k}^{i}) = \log\frac{p(b_{k}^{i} = 1|y_{n})}{p(b_{k}^{i} = 0|y_{n})},
\end{equation}
where $b_{k}^{i}$ is the $i$th bit of layer $k$ and the subscript $1$ denotes the LLR information associated with SCMA decoder. By Bayes' law,
\begin{align}\label{06b}
  \nonumber L_{1}(b_{k}^{i}) & = \log\frac{p(y_{n}|b_{k}^{i} = 1)}{p(y_{n}|b_{k}^{i} = 0)} + \log\frac{p(b_{k}^{i} = 1)}{p(b_{k}^{i} = 0)} \\
   & = L^{e}_{1} + L^{a}_{1},
\end{align}
where $L^{e}_{1}$ and $L^{a}_{1}$ are the extrinsic and \emph{a priori} LLR of SCMA decoder, respectively. Similarly, for the channel decoder, we can also have
\begin{equation}\label{06c}
  L_{2}(b_{k}^{i}) = L^{e}_{2} + L^{a}_{2},
\end{equation}
where the subscript $2$ denotes the LLR information associated with channel decoder.

Thus, both SCMA decoder and channel decoder can be viewed as a box that outputs the extrinsic LLR information given the \emph{a priori} LLR information computed from the other as well as the observations from the radio channel. The two boxes work cooperatively to exchange the extrinsic LLR information until a preset maximum number of iterations is reached.

The sparsity of SCMA codewords makes the near optimal message passing algorithm possible. As a belief propagation based decoding algorithm, in MPA, the soft information of SCMA codewords are iteratively exchanged between layer nodes and resource nodes. In~\cite{28}, several graphical models including Markov random fields, Tanner graph and Bayesian networks for MPA are introduced. In addition,~\cite{28} has also discussed some different marginalization algorithms in MPA.

In log domain, the message sent from resource node $n$ to layer node $k$ is given by,
\begin{equation}\label{07}
  I_{g_{n}\rightarrow v_{k}}(\mathbf{x}_{k}) = \max^{*}_{\mathbf{x}_{u}:u\in\xi_{n}\setminus k}\{-f_{n}(\mathbf{x})+\sum_{u\in\xi_{n}\setminus k}I_{v_{u}\rightarrow g_{n}}(\mathbf{x}_{u})\},
\end{equation}
where $f_{n}(\mathbf{x}) = \frac{1}{\sigma^{2}}\|y_{n}-\Sigma_{k\in\xi_{n}}h_{k,n}x_{k,n}\|^{2}$ with $y_{n}$ being the received signal in subcarrier $n$ and $x_{k,n}$ being the $n$th component of the SCMA codeword $\mathbf{x}_{k}$. The $\max\limits^{*}$ operation is given by
\begin{align}\label{08}
  \nonumber \max\limits^{*}(a,b) & = \log(e^{a}+e^{b})  \\
                                 & = \max(a,b) + \log(1+e^{-|a-b|}).
\end{align}
To facilitate hardware implementation, Max-log-MPA is often used and we have the approximation $\max\limits^{*}(a,b) \approx \max(a,b)$.

The message sent from layer node $k$ to resource node $n$ is given by,
\begin{equation}\label{09}
I_{v_{k}\rightarrow g_{n}}(\mathbf{x}_{k}) = L(\mathbf{x}_{k}) + \sum_{l\in \zeta_{k}\setminus n} I_{g_{l}\rightarrow v_{k}}(\mathbf{x}_{k}),
\end{equation}
where $ L(\mathbf{x}) = \log p(\mathbf{x})$ is \emph{a priori} probability of $\mathbf{x}$ received from the channel decoder. When the algorithm is converged or a maximum number of iterations is reached, a \emph{posteriori} probability of codeword $\mathbf{x}_{k}$ is given by,
\begin{equation}\label{10}
  I(\mathbf{x}_{k}) = L(\mathbf{x}_{k}) + \sum_{l\in \zeta_{k}} I_{g_{l}\rightarrow v_{k}}(\mathbf{x}_{k}).
\end{equation}

The main cost of the MPA algorithm lies in~\eqref{07}, where to get the LLR of $\mathbf{x}_{k}$ we need to marginalize over $M^{d_{c}-1}$ hypotheses. Since the constellation size is $M$, the total complexity of~\eqref{07} is $M^{d_{c}}$ in each iteration. Clearly the marginalization problem has the exponential complexity and is computational intractable when the scale of problem is large.

To reduce the decoding complexity, in this paper, we propose a low complexity algorithm by exploring the lattice structure of SCMA codewords. Before introducing our algorithm, we rewrite the received signal on subcarrier $n$ as,
\begin{align}\label{11}
  \nonumber y_{n} & = \sum_{k\in\xi_{n}}h_{n,k}x_{n,k}+z \\
  \nonumber        & = \mathbf{h}^{T}_{n}\mathbf{\bar{x}}_{n}+z \\
  \nonumber        & = \mathbf{h}^{T}_{n}\mbox{diag}\{\mathbf{g}^{T}_{n,\kappa(1)},\ldots,\mathbf{g}^{T}_{n,\kappa(d_{c})}\}
                     \mathbf{u}_{n}+z \\
          & = \mathbf{H}_{n}\mathbf{u}_{n}+z,
\end{align}
where $\mathbf{\bar{x}}_{n}=[x_{n,\kappa(1)}, \ldots, x_{n,\kappa(d_{c})}]^{T}$ are the set of collision users in the $n$th subcarrier, $\mathbf{h}_{n}=[ h_{n,\kappa(1)}, \ldots, h_{n,\kappa(d_{c})} ]^{T}$ are their channel coefficients corresponding to subcarrier $n$, $z$ is the complex Gaussian noise. Since $\mathbf{x}_{P,k}=\mathbf{G}_{k}\mathbf{u}_{2P,k}$, the $n$th component of $\mathbf{x}_{P,k}$ can be written as $x_{n,k} = \mathbf{g}_{n,k}^{T}\mathbf{u}_{2P,k}$, where row vector $\mathbf{g}_{n,k}^{T}$ corresponds to the $n$th row of matrix $\mathbf{G}_{k}$. Therefore, we can write $\mathbf{\bar{x}}_{n} = \mbox{diag}\{\mathbf{g}^{T}_{n,\kappa(1)},\ldots,\mathbf{g}^{T}_{n,\kappa(d_{c})}\}\mathbf{u}_{n}$ with $\mathbf{u}_{n}=[ \mathbf{u}_{2P,\kappa(1)}^{T},\ldots,\mathbf{u}_{2P,\kappa(d_{c})}^{T} ]^{T}$ being the set of lattice points of collision users in subcarrier $n$. Let $\mathbf{H}_{n} = \mathbf{h}^{T}_{n}\mbox{diag}\{\mathbf{g}^{T}_{n,\kappa(1)},\ldots,\mathbf{g}^{T}_{n,\kappa(d_{c})}\}$, equation~\eqref{11} can be obtained. From the discussion above, $f_{n}(\mathbf{x})$ in~\eqref{07} can be reformulated as $f_{n}(\mathbf{u}_{n})= \frac{1}{\sigma^{2}}\|y_{n}-\mathbf{H}_{n}\mathbf{u}_{n}\|^{2}$.

\section{Low Complexity SCMA Decoder}
In this section, we propose a new detection algorithm which is based on list sphere decoding. In MPA, to obtain~\eqref{07} an exponential number of Euclidean distances $f_{n}(\mathbf{x})$ are required to compute. We can observe that this massive computation is useless since for some Euclidean distances that are large enough, $e^{-f_{n}(\mathbf{x})}\approx 0$ and thus have tiny contributions to the value of~\eqref{07}. The main idea of sphere decoding is that instead of the exhaustive search for all possible hypotheses, only signals within a given hypersphere are considered.

\subsection{Sphere Decoder for SCMA}
Consider an ML receiver,
\begin{equation}\label{12}
  \mathbf{\hat{u}}_{n} = \arg\min_{\mathbf{u}_{n} \in \Lambda} \| y_{n}- \mathbf{H}_{n}\mathbf{u}_{n}\|^{2},
\end{equation}
where $y_{n}$ is the received signal in the $n$th subcarrier, $\mathbf{H}_{n}$ and $\mathbf{u}_{n}$ are respectively row and column vectors defined as in section III-B. We assume here that the size of the two vectors is $L$.

In original sphere decoder discussed in~\cite{16,17,18,19}, $\mathbf{H}_{n}$ is an $N\times M$ matrix with $N\geq M$ (e.g., the number of receive antennas is required to be no less than that of transmit antennas in MIMO systems), this makes the cholesky decomposition of the matrix $\mathbf{H}_{n}^{T}\mathbf{H}_{n}$ tractable. In the formulation of equation~\eqref{11}, however, $\mathbf{H}_{n}$ is a row vector and the system is therefore underdetermined. In order to implement the QR factorization to the vector $\mathbf{H}_{n}$, we reformulate equation~\eqref{12} as follows~\cite{33},
\begin{equation}\label{13}
  \mathbf{\hat{u}}_{n} = \arg\min_{\mathbf{u}_{n} \in \Lambda} \| y_{n}- \mathbf{H}_{n}\mathbf{u}_{n}\|^{2} + \alpha
  \mathbf{u_{n}}^{\dag}\mathbf{u_{n}},
\end{equation}
where the entries of $\mathbf{u}_{n}$ are assumed to be constant modulus (e.g., BPSK) and $\alpha > 0$. Rewrite~\eqref{13} as
\begin{align}\label{14}
    \nonumber  \mathbf{\hat{u}}_{n}& = \arg\min_{\mathbf{u}_{n} \in \Lambda} \| y_{n}- \mathbf{H}_{n}\mathbf{u}_{n}\|^{2} + \alpha\mathbf{u_{n}}^{\dag}\mathbf{u_{n}}\\
      & = \arg\min_{\mathbf{u}_{n} \in \Lambda}\left \|\mathbf{\widetilde{y}}_{n} - \mathbf{\widetilde{H}}_{n}\mathbf{u}_{n}
      \right\|^{2},
\end{align}
where $\mathbf{\widetilde{y}}_{n} = \begin{pmatrix}
                                      y_{n} \\
                                      \mathbf{0}  \\
                                    \end{pmatrix} $ and $\mathbf{\widetilde{H}}_{n} = \begin{pmatrix}
                                                           \mathbf{H}_{n} \\
                                                           \alpha \mathbf{I} \\
                                                         \end{pmatrix}$ is $(L+1) \times 1$ vector and $(L+1) \times L$ matrix, respectively.

Now matrix $\mathbf{\widetilde{H}}_{n}$ is a column full rank matrix, apply the QR factorization
\begin{equation}\label{15}
  \mathbf{\widetilde{H}}_{n} = \left[\mathbf{Q}_{1}, \mathbf{Q}_{2}\right]\begin{bmatrix}
                                                     \mathbf{R} \\
                                                     \mathbf{0} \\
                                                   \end{bmatrix},
\end{equation}
where the dimension of $\mathbf{Q}_{1}$, $\mathbf{Q}_{2}$ and $\mathbf{R}$ are $(L+1)\times L$, $(L+1) \times 1$ and $L\times L$, respectively. Since multiplication of unitary matrix does not change the norm of vectors, equation~\eqref{14} can be written as,
\begin{align}\label{16}
    \nonumber  \mathbf{\hat{u}}_{n} & = \arg\min_{\mathbf{u}_{n} \in \Lambda}\left \|\mathbf{\widetilde{y}}_{n} - \mathbf{\widetilde{H}}_{n}\mathbf{u}_{n} \right\|^{2} \\
     & = \arg\min_{\mathbf{u}_{n} \in \Lambda}\left \|\mathbf{y'}_{n} - \mathbf{R}\mathbf{u}_{n} \right\|^{2},
\end{align}
where $\mathbf{y'}_{n} = \mathbf{Q}_{1}^{\dag}\mathbf{\widetilde{y}}_{n}$. For the sake of notation simplification, we will drop the prime of $\mathbf{y'}_{n}$ in the following.

In sphere decoding, instead of exhaustive search for all possible hypotheses, only signals within a given radius $C$ are considered, i.e.,
\begin{equation}\label{17}
  \left\|\mathbf{y}_{n} - \mathbf{R}\mathbf{u}_{n} \right\|^{2} \leq C.
\end{equation}
Since $\mathbf{R}$ is an upper triangular matrix, we can expand inequality~\eqref{17} as,
\begin{equation}\label{18}
  \sum_{i=1}^{L} |y_{i} - \sum_{j=i}^{L}r_{ij}u_{j}|^{2} \leq C.
\end{equation}
The sphere decoder works in a backward recursive way to find the entries of $u_{n}$ that are lied in a hypersphere with radius $C$.

To start, consider the last term in the left hand side of~\eqref{18}. Let $i=L$, $T_{L} = C$, $\hat{y}_{L} = y_{L}$. We have
\begin{equation}\label{19}
  |\hat{y}_{L} - r_{LL}u_{L}|^{2} \leq T_{L}.
\end{equation}
For complex $\hat{y}_{L}$, let $\hat{y}_{L}=|\hat{y}_{L}|\exp(\hat{\theta}_{L})$. Then inequality~\eqref{19} can be expanded as,
\begin{equation}\label{20}
  r_{LL}^{2}u_{L}^{2} - 2|\hat{y}_{L}|\cos(\hat{\theta}_{L})r_{LL}u_{L}+|\hat{y}_{L}|^{2}-T_{L}\leq 0.
\end{equation}
Solving the above quadratic inequality, we can get the upper bound of $u_{L}$
\begin{equation}\label{21}
  UB(u_{L}) = \frac{|\hat{y}_{L}|\cos(\hat{\theta}_{L})+\sqrt{T_{L}-|\hat{y}_{L}|^{2}\sin^{2}(\hat{\theta}_{L})}}{r_{LL}},
\end{equation}
and the lower bound of $u_{L}$
\begin{equation}\label{22}
  LB(u_{L}) = \frac{|\hat{y}_{L}|\cos(\hat{\theta}_{L})-\sqrt{T_{L}-|\hat{y}_{L}|^{2}\sin^{2}(\hat{\theta}_{L})}}{r_{LL}}.
\end{equation}
So we have $LB(u _{L})\leq u _{L}\leq UB(u _{L})$.

In general, for $i=l$, we have
\begin{align}\label{23}
  \nonumber |y_{l} - &\sum_{j=l+1}^{L}r_{lj}u_{j} - r_{ll}u_{ll}|^{2} \\
   & + \sum_{i=l+1}^{L}|y_{i} - \sum_{j=i}^{L}r_{ij}u_{j}|^{2} \leq C.
\end{align}
Let $\hat{y}_{l} = y_{l} - \sum_{j=l+1}^{L}r_{ij}u_{j}$ and $T_{l} = C - \sum_{i=l+1}^{L}|\hat{y}_{i}-r_{ii}u_{i}|^{2}$, inequality~\eqref{23} can be simplified as,
\begin{equation}\label{24}
  |\hat{y}_{l} - r_{ll}u_{l}|^{2} \leq T_{l}.
\end{equation}
Clearly, we can get the upper and lower bound of $u_{l}$ in a same manner as in~\eqref{19} so that
\begin{equation}\label{25}
  UB(u_{l}) = \frac{|\hat{y}_{l}|\cos(\hat{\theta}_{l})+\sqrt{T_{l}-|\hat{y}_{l}|^{2}\sin^{2}(\hat{\theta}_{l})}}{r_{ll}},
\end{equation}
and
\begin{equation}\label{26}
  LB(u_{l}) = \frac{|\hat{y}_{l}|\cos(\hat{\theta}_{l})-\sqrt{T_{l}-|\hat{y}_{l}|^{2}\sin^{2}(\hat{\theta}_{l})}}{r_{ll}}.
\end{equation}
The algorithm runs in a backward recursive way until $i=1$ and $LB(u_{1})\leq u_{1}\leq UB(u_{1})$. In this case, a point within the sphere has been found and we have,
\begin{equation}\label{27}
   \left\|\mathbf{y}_{n} - \mathbf{R}\mathbf{u}_{n} \right\|^{2} = T_{L} - T_{1} + |y_{1} - r_{11}u_{1}|^{2}.
\end{equation}

During the search in each level $l$, we can set $u_{l} = \lceil LB(u_{l})\rceil-1$ and proceed in a lexicographic order until it reaches $\lfloor UB(u_{l})\rfloor$. This is the original version of sphere decoder proposed in~\cite{14}. Schnorr-Euchner (SE) enumeration~\cite{27} is another strategy where the possible points are sorted in an ascending order according to their distance increment (DI) $e(u_{l})=|\hat{y}_{l} - r_{ll}u_{l}|^{2}$, i.e., signals are searched in the order,
\begin{equation}\label{27a}
  u_{l,1},u_{l,2},u_{l,3},... ,
\end{equation}
such that $e(u_{l,i})\leq e(u_{l,j})$ for $i<j$. When a point $u_{l,i}$ failed to satisfy inequality~\eqref{23}, the point $u_{l,j}$ with $i<j$ would also fail. Therefore, the search for the rest candidate point can be skipped. The SE enumeration can find the right path earlier than the original sphere decoder since in each time we begin with the point that minimizes the partial Euclidean distance (PED) $d(\mathbf{u}^{L}_{l})=\sum_{i=l}^{L}|y_{i}-\sum_{j=i}^{L}r_{ij}u_{j}|^{2}$.

To render the search in ascending order according to the DI in each level, we let  $u_{l}=\mbox{sign}(|\hat{y_{l}}|\cos(\hat{\theta}_{l}))$ and set the search step as $\Delta_{l}=-2\cdot\mbox{sign}(|\hat{y_{l}}|\cos(\hat{\theta}_{l}))$. Clearly, the calculation of $LB(u_{l})$ is avoided with the above settings. Thus, the search is free of the initial radius and can be set as $C=+\infty$ in SE enumeration.

In the above discussion, we have assumed that the elements in $\mathbf{u}_{n}$ are of constant modulus. However, for QAM symbols, the real and imaginary part are two pulse amplitude modulation (PAM) sets, i.e., $u_{i} = \pm1, \pm3, \cdots$. In this case,~\eqref{13} may never be equivalent to~\eqref{12}. To solve this problem, a simple decomposition of the entries in $\mathbf{u}_{n}$ can be carried out. Observe that for $u\in$ M-PAM and $u_{i}' \in$ BPSK, we have,
\begin{equation}\label{28}
  u = \sum_{i=0}^{\log M-1}2^{i}\cdot u_{i}'=\pmb{\gamma}^{T}  \cdot \mathbf{u}',
\end{equation}
where $\pmb{\gamma}^{T} = [2^{\log M-1}, \ldots , 2 , 1]$. Hence, the $2P$-dimensional non-zero signal $\mathbf{u}_{2P,k}$ for user $k$ can be written as,
\begin{equation}\label{29}
   \mathbf{u}_{2P,k}= \begin{bmatrix}
                        \pmb{\gamma}^{T}_{1,k}  &   &   &   \\
                          & \pmb{\gamma}^{T}_{2,k}  &   &   \\
                          &   & \ddots &   \\
                          &  &   & \pmb{\gamma}^{T}_{2P,k}  \\
                      \end{bmatrix}
                      \begin{bmatrix}
                        \mathbf{u}_{k}^{'1} \\
                        \mathbf{u}_{k}^{'2} \\
                        \ldots \\
                        \mathbf{u}_{k}^{'2P} \\
                      \end{bmatrix} =\pmb{\Gamma}_{k}\cdot \mathbf{u}_{2P,k}',
\end{equation}
with the entries of $\mathbf{u}_{2P,k}'$ being BPSK signals. In addition, the signals in subcarrier $n$ can be written as,
\begin{align}\label{30}
   \nonumber \mathbf{u}_{n} &= \begin{bmatrix}
                                 \mathbf{u}_{2P,\kappa(1)} \\
                                 \mathbf{u}_{2P,\kappa(2)} \\
                                 \ldots \\
                                 \mathbf{u}_{2P,\kappa(d_{c})} \\
                                \end{bmatrix} \\
                            &= \begin{bmatrix}
                                 \pmb{\Gamma}_{\kappa(1)}&   &   &   \\
                                   & \pmb{\Gamma}_{\kappa(2)} &   &   \\
                                   &   & \ddots &   \\
                                   &   &   & \pmb{\Gamma}_{\kappa(d_{c})} \\
                               \end{bmatrix} \begin{bmatrix}
                                               \mathbf{u}_{2P,\kappa(1)}' \\
                                               \mathbf{u}_{2P,\kappa(2)}' \\
                                               \ldots \\
                                               \mathbf{u}_{2P,\kappa(d_{c})}' \\
                                             \end{bmatrix}.
\end{align}

With the decomposition of~\eqref{30}, the received signal on the $n$th subcarrier can be rewritten as,
\begin{align}\label{31}
  \nonumber y_{n} &= \sum_{k\in\xi_{n}}h_{n,k}x_{n,k}+z \\
  \nonumber       &= \mathbf{h}^{T}_{n}\mathbf{x}_{n}+z \\
  \nonumber       &= \mathbf{h}^{T}_{n}\mbox{diag}\{\mathbf{g}^{T}_{n,\kappa(1)},\ldots,\mathbf{g}^{T}_{n,\kappa(d_{c})}\}
                      \mathbf{u}_{n}+z \\
  \nonumber       &= \mathbf{h}^{T}_{n}\mbox{diag}\{\mathbf{g}^{T}_{n,\kappa(1)}\pmb{\Gamma}_{\kappa(1)},
                      \ldots,\mathbf{g}^{T}_{n,\kappa(d_{c})}\pmb{\Gamma}_{\kappa(d_{c})}\}
                      \mathbf{u}_{n}'+z \\
                  &= \mathbf{H} _{n}'\mathbf{u}_{n}'+z,
\end{align}
where vector $\mathbf{H}_{n}' = \mathbf{h}^{T}_{n}\mbox{diag}\{\mathbf{g}^{T}_{n,\kappa(1)}\pmb{\Gamma}_{\kappa(1)}, \ldots,\mathbf{g}^{T}_{n,\kappa(d_{c})}\pmb{\Gamma}_{\kappa(d_{c})}\}$ and $\mathbf{u}_{n}'=[ \mathbf{u}_{2P,\kappa(1)}^{'T},\ldots,\mathbf{u}_{2P,\kappa(d_{c})}^{'T} ]^{T}$. In this way, the decoding algorithm can run in the same manner as the constant modulus case.
\begin{algorithm}[t]
    \caption{LSD based MPA Detection}
    \label{alg1}
    \begin{algorithmic}[1]
    \REQUIRE $\mathbf{Q}=\left[\mathbf{Q}_{1}, \mathbf{Q}_{2}\right],\mathbf{R},\mathbf{y}_{n} = \mathbf{Q}^{\dag}_{1}\mathbf{\widetilde{y}}_{n}, C=+\infty, IT$
    \ENSURE (LSD Iteration)
        \STATE Set $i=L, \hat{y}_{L}=y_{L}, T_{L}=0$
        \STATE (\emph{SE Enumeration}) Set $u_{i}=\mbox{sign}(|\hat{y_{i}}|\cos(\hat{\theta}_{i}))$ and search step $\Delta_{i}=-2\cdot\mbox{sign}(|\hat{y_{i}}|\cos(\hat{\theta}_{i}))$.
        \STATE (\emph{Node Pruning}) If $T_{i}+|\hat{y}_{i}-r_{ii}u_{i}|^{2} > C$ or $u_{i}\notin\{-1,1\}$, go to 4.
               else go to 5.
        \STATE If $i=L$, terminate and output $\Phi_{n}$; else $i=i+1$, $u_{i}=u_{i}+\Delta_{i}$, go to 3.
        \STATE (\emph{PED Computation}) If $i=1$, go to 6; else set $T_{i-1}=T_{i}+|\hat{y}_{i}-r_{ii}u_{i}|^{2}$, $\hat{y}_{i-1}=y_{i-1}-\sum_{j=i}^{L}r_{i-1j}u_{j}$ and $i=i-1$, go to 2.
        \STATE A lattice point $\mathbf{u}_{n}$ within the hypersphere has been found. Let $\left\|\mathbf{y}_{n} - \mathbf{R}\mathbf{u}_{n} \right\|^{2} = T_{1} + |y_{1} - r_{11}u_{1}|^{2}$. Add $\mathbf{u}_{n}$ and the corresponding Euclidean distance $d(\mathbf{u}_{n})$ to the candidate list set. If the candidate list set is full (the size reaches $T_{max}$), find $\mathbf{u}_{n}$ with maximum $d(\mathbf{u}_{n})$, set $C=d(\mathbf{u}_{n})$. $u_{i}=u_{i}+\Delta_{i}$, go to 3.
    \ENSURE (MPA Iteration)
       \STATE Set $i=1$
       \WHILE{$i\leq IT$}
         \STATE Compute $I_{g_{n}\rightarrow v_{k}}(\mathbf{x}_{k})$ using~\eqref{32}.
         \STATE Compute $I_{v_{k}\rightarrow g_{n}}(\mathbf{x}_{k})$ using~\eqref{09}.
         \STATE $i=i+1$.
       \ENDWHILE
       \STATE (\emph{LLR of} $\mathbf{x}_{k}$) Compute $I(\mathbf{x}_{k})$ using~\eqref{10}.
    \end{algorithmic}
\end{algorithm}

\subsection{List Sphere Decoding based MPA}
The list sphere decoder will return all the candidate lattice points within a given radius, i.e., all $\mathbf{u}_{n}$ satisfying~\eqref{17}. Now let the candidate list of lattice points be $\Phi_{n} = \{\mathbf{u}_{n}^{1},\ldots,\mathbf{u}_{n}^{T_{max}}\}$, where $T_{max}$ is the size of the candidate list set. Equation~\eqref{07} can be approximated as,
\begin{align}\label{32}
  \nonumber I_{g_{n}\rightarrow v_{k}}&(\mathbf{x}_{k}=\mathbf{x}_{k}^{m})\\
  & \approx  \max^{*}_{\mathbf{x}_{u} \in \Phi_{n}\cap \mathbb{X}_{k,n}^{m} }\{-f_{n}(\mathbf{x}) +\sum_{u\in\xi_{n}\setminus k}I_{v_{u}\rightarrow g_{n}}(\mathbf{x}_{u})\},
\end{align}
where $\mathbb{X}_{k,n}^{m}$ is the set of codewords collided on subcarrier $n$ with $\mathbf{x}_{k}$ being the $m$th constellation point.

The idea behind this approximation is that signals which have large Euclidean distances from received value $y_{n}$ actually have tiny contribution to~\eqref{32}. Hence they are excluded in the computation $\max\limits^{*}(\cdot)$ so as to reduce the decoding complexity. Clearly, the larger $T_{max}$, the more precise the approximation~\eqref{32} and the equality holds when $T_{max}=M^{d_{c}}$. However, small $T_{max}$ will result in a low computational complexity. So there is a trade off between the performance and decoding complexity. As we can observe in Section V, one can achieve a negligible performance loss with a $T_{max} \ll M^{dc}$.

We will refer the algorithm LSD-MPA for short and summarize it in Algorithm~\ref{alg1}.

\subsection{Further Complexity Reductions}
\begin{figure}
  \centering
  \includegraphics[width=3.5in]{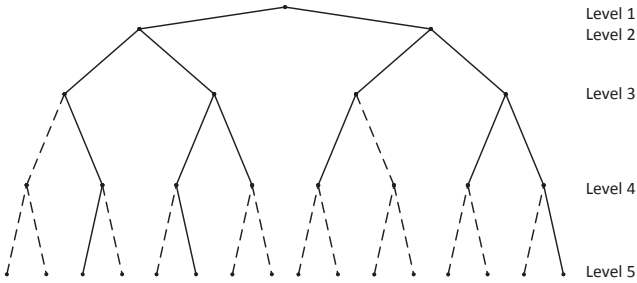}\\
  \caption{Sphere decoder: depth-first tree search}\label{Fig.3}
\end{figure}

Sphere decoder can be viewed as a kind of depth-first tree search algorithm. Since the entries of $\mathbf{u}_{n}'$ are decomposed to have BPSK signals, the depth of the search tree is given by $L=d_{c}\cdot \log_{2}M$. In Fig.~\ref{Fig.3}, we plot a search tree with $L=5$ as an example, where the search process starts from the root node and proceeds to the leaf nodes. All solid lines correspond to the survival pathes during the search while the dotted lines correspond to the eliminated pathes. In general, the complexity of sphere decoder depends on the size of the search tree: the smaller the number of nodes visited during the search, the lower the complexity of the algorithm~\cite{30}. Thus, to reduce the complexity of sphere decoder, we should manage to reduce the total visited nodes during the tree search process.

First we shall discuss the choice of initial search radius. In SE enumeration, the search radius can be set as $C=+\infty$ which guarantees that we can always find a point within the hypersphere. The first point searched by SE enumeration is referred to Babai point~\cite{18}. In sphere decoding, once a new point has been found, the search radius is updated to the corresponding value of that point. However, in LSD, since what we want is a candidate list set instead of a single point, the radius cannot be updated until the candidate list set is full (the size reaches $T_{max}$). In other words, the radius will remain to be infinity until the $T_{max}$th point has been found. Therefore, too many redundancy nodes will be visited during the search since the pruning condition in step 3 of Algorithm~\ref{alg1} failed to work until $C$ is updated.

To make the pruning condition work before the candidate list set is full, we should determine a proper initial search radius that is not only tighten enough to prune the redundancy nodes but also large enough to make the candidate list set effective.

Let $d(\mathbf{u}_{n}) = \| y_{n}- \mathbf{H}_{n}\mathbf{u}_{n}\|^{2} + \alpha
  \mathbf{u_{n}}^{\dag}\mathbf{u_{n}}$. We choose $C$ such that
\begin{equation}\label{32a}
  P\{C < d(\mathbf{u}_{n})\} < \epsilon,
\end{equation}
where $\epsilon$ is a small enough number, i.e., the search radius is almost surely larger than $d(\mathbf{u}_{n})$. When decoding is perfect, $d(\mathbf{u}_{n}) = \|z_{n}\|^{2} + \alpha L$. Thus we can have,
\begin{align}\label{32b}
   \nonumber P\{C < d(\mathbf{u}_{n})\} & = P\{C < \|z_{n}\|^{2} + \alpha L\} \\
   \nonumber & = P\left\{\frac{2(C - \alpha L)}{\sigma^{2}} < \frac{2\|z_{n}\|^{2}}{\sigma^{2}}\right\} \\
   & = 1 - F\left[\frac{2(C - \alpha L)}{\sigma^{2}};2\right],
\end{align}
where $\frac{2\|z_{n}\|^{2}}{\sigma^{2}}$ is a chi-square random variable with 2 degrees of freedom since $z_{n}\sim\mathcal{CN}(0, \sigma^{2})$ and $F(x;k)$ is the cumulative distribution function of chi-square variable $x$ with $k$ degrees of freedom. From~\eqref{32b}, we can determine the initial search radius as,
\begin{equation}\label{32c}
  C = \frac{\sigma^{2}}{2} F^{-1}(1-\epsilon;2)+\alpha L.
\end{equation}

Another fact that should be noted lies in the approximation of equation~\eqref{32}. In LSD based MPA, we approximate~\eqref{32} in subcarrier $n$ by using a candidate list set $\Phi_{n}$ whose size is far less than the whole search space. Denote $\mathbf{x}_{k}^{m}$ as the $m$th constellation signal point of layer $k$. When $\mathbf{x}_{k}^{m}$ is contained in $\Phi_{n}$, $I_{g_{n}\rightarrow v_{k}}(\mathbf{x}_{k}^{m})$ is calculated using~\eqref{32}. On the contrary, when $\mathbf{x}_{k}^{m}$ is excluded in $\Phi_{n}$, we set the corresponding LLR to an extreme value. The reason is that when no $\mathbf{x}_{k}^{m}$ is contained in $\Phi_{n}$, the possibility of $\mathbf{x}_{k}^{m}$ being the actual transmit signal is small. Thus in log domain, we set $I_{g_{n}\rightarrow v_{k}}(\mathbf{x}_{k}^{m}) = -\infty$.

Now consider~\eqref{09}, i.e., the message sent from layer $k$ to resource node $n'\in \zeta_{k} \setminus n$. When $I_{g_{n}\rightarrow v_{k}}(\mathbf{x}_{k}^{m}) = -\infty$, we can immediately get $I_{v_{k}\rightarrow g_{n'}}(\mathbf{x}_{k}^{m}) = -\infty$. In this case, when computing~\eqref{32} in subcarrier $n'$, $\mathbf{x}_{k}^{m}$ would have no contribution to~\eqref{32} either for $\max^{*}(a,b)$ or $\max(a,b)$. Thus in candidate points search in subcarrier $n'$, when $\mathbf{x}_{k}^{m}$ is found, we just drop this point and skip the search for the remaining pathes. By doing so, the efforts for candidate search could be saved considerably. For example, in the factor graph showed in Fig.~\ref{Fig.2}, after conducting the LSD in subcarrier g$4$, we find $\mathbf{x}_{6}^{m}$, the $m$th constellation point for layer $6$ is not contained in the candidate list set. Next, in the search on subcarrier g$3$, when layer $6$ is found to be $\mathbf{x}_{1}^{m}$, we skip the remaining search for layers $2$ and $4$ since no matter which constellation point they choose, it has no effects on the final results on~\eqref{32}. Note that user $6$ is decoded first in each subcarrier since the LSD works in a backward way and for the same reason we start the decoding from the last subcarrier.

Since the nodes being pruned has no effects on~\eqref{32}, the accuracy of the candidate list set remains unaffected. What's more, the performance of LSD-MPA may be even improved. The reason is as follows. In LSD, an error occurs when the actual transmit signal $\mathbf{u}_{n}^{t}$ is not contained in the candidate list set $\Phi_{n}$. This would happen when $d(\mathbf{u}_{n}^{t})$ is larger than the maximum value in $\Phi_{n}$. With the pruning of redundancy nodes that do not affect the accuracy of $\Phi_{n}$, it is likely that $\mathbf{u}_{n}^{t}$ would enter the final candidate list set in this case and thus improve the performance of LSD-MPA. The reasoning will be confirmed by the simulation results in Section V.

For LNP SCMA, there is a number of overlapped points in each OFDMA subcarrier. As in~\cite{09b}, we define $T_{k,n}$ as the set of indices corresponding to the repetitive projections points for layer $k$ over subcarrier $n$. For $i,j \in T_{k,n}$, we have $I_{g_{n}\rightarrow v_{k}}(\mathbf{x}_{k}^{i}) = I_{g_{n}\rightarrow v_{k}}(\mathbf{x}_{k}^{j})$ where $\mathbf{x}_{k}^{i}$ denotes the $i$th constellation point for layer $k$. In the search tree shown in Fig~\ref{Fig.3}, this corresponds to two branches that have the same metric. Therefore we would retain only one branch of the tree and prune the others that have the same metrics. By doing so, a small size of the search tree has been obtained.

In the following we will refer to the LSD based MPA with node prunning as NP-LSD-MPA in short.

\begin{table*}[ht]
  \centering
  \caption{Complexity Comparison of Different Decoding Algorithms}\label{I}
  \begin{tabular}{|c||c|c|c|c|}
  \hline
              & Sphere Decoding & $I_{g_{n}\rightarrow v_{k}}(\mathbf{x}_{k}) $ & $I_{v_{k}\rightarrow g_{n}}(\mathbf{x}_{k})$
              & $I(\mathbf{x}_{k})$ \\ [0.5ex] \hline\hline
  MPA 　　　　& - & $IT\cdot N(3d_{c}^{2}+3d_{c}) M^{d_{c}}$
              & $IT\cdot N(\frac{N}{K}d_{c}^{2}-d_{c}) M$ & $Nd_{c}M$ \\ \hline
  PM-MPA　    & - &$mN(3d_{c}^{2}+3d_{c})M^{d_{c}}+(IT-m)(3d_{c}^{2}+3d_{c})C_{PM}$
              & $IT\cdot N(\frac{N}{K}d_{c}^{2}-d_{c})M$
              & $Nd_{c}M$ \\ \hline
  LSD-MPA　   & \tabincell{c}{$N[2L^{3}+2L^{2}+L$  \\ $+\sum_{k=1}^{L}(2k+7)N_{k}]$}
              & $IT\cdot Nd_{c}^{2}T_{max}+N\widehat{N}_{L}T_{max}$  & $IT\cdot N(\frac{N}{K}d_{c}^{2}-d_{c})M$
              & $Nd_{c}M$ \\ \hline
  \end{tabular}
\end{table*}

\subsection{Complexity Analysis}
The complexity of the decoding algorithms proposed in the previous section is analysed here. It is assumed that a floating-point (FLOP) operation is either a complex multiplication or a complex summation in this paper. Note that a complex multiplication needs six real operations while a complex summation needs two real operations. For SCMA decoder, Max-log-MPA is used.

First we analyse the complexity of sphere decoding. Before running the algorithm, we need to conduct QR factorization of matrix $\widetilde{\mathbf{H}}_{n}$ and compute $\mathbf{y'}_{n} = \mathbf{Q}_{1}^{\dag}\mathbf{\widetilde{y}}_{n}$. Note that we have the decomposition~\eqref{31}, so $\mathbf{H}_{n}$ is an $1\times L$ row vector with $L=d_{c}\cdot\log_{2}M$. For QR factorization, we use the modified Gram-Schmidt (MGS) algorithm~\cite{34}, which requires $2(L+1)L^{2}$ flops. Vector computation of $\mathbf{y'}_{n}$ requires $L$ flops. Therefore, $2L^{3}+2L^{2}+L$ FLOPs in total are needed before running sphere decoding.

For sphere decoding, expected complexity~\cite{31,32} as showed in equation~\eqref{33} is considered in this paper,
\begin{equation}\label{33}
 E(k)  = \sum_{k=1}^{L}f(k)N_{k} = \sum_{k=1}^{L}(2k+7)N_{k},
\end{equation}
where $N_{k}$ and $f(k)$ are the average number of visited nodes during the tree search and the number of elementary arithmetic operations needed (from step 2 to step 6 in Algorithm~\ref{alg1}) in level $k$, respectively. Thus, the complexity of sphere decoding depends mainly on the expected number of visited node in each level of the search tree.

Next, we consider the MPA decoding. For Max-log-MPA, the computation of~\eqref{07} requires $N(2d_{c}^{2}-d_{c})M^{d_{c}}$ summations and $N(d_{c}^{2}+3d_{c})M^{d_{c}}$ multiplications whereas for LSD based MPA, $N(d_{c}^{2}-d_{c})T_{max}$ summations are needed since $f_{n}(\mathbf{x})$ has been computed in Algorithm~\ref{alg1}. Further, $Nd_{c}(P-1)M = (\frac{N^{2}}{K}d_{c}^{2}-Nd_{c})M$ summations are needed for~\eqref{09}, and~\eqref{10} requires $KPM=Nd_{c}M$ summations. Notice that when using the approximation $\max\limits^{*}(a,b) \approx \max(a,b)$, we need $Nd_{c}M^{d_{c}}$ comparisons for Max-log-MPA. For LSD based MPA, the number has been reduced to $Nd_{c}T_{max}$ comparisons. In addition, in step $6$ of Algorithm~\ref{alg1}, to find the maximum $d(\mathbf{u}_{n})$ that updates the radius $C$, we need $T_{max}$ comparisons in general and the total comparison in LSD is therefore $N\widehat{N}_{L}T_{max}$, where $\widehat{N}_{L}$ is less than $N_{L}$, the number of visited nodes in the root of the search tree since the radius is only updated when $d(\mathbf{u}_{n}) < C$. Otherwise $C$ remains unchanged and the search is unnecessary.

For PM based MPA, the complexity analysis is similar to that in~\cite{11}. In the first $m$ iterations, the complexity is the same to MPA since no reference symbols are chosen. After the $m$th iteration, the computational complexity in the resource node is reduced to $C_{PM}(2d_{c}^{2}-d_{c})$ summations and $C_{PM}(d_{c}^{2}+3d_{c})$ multiplications where $C_{PM}=(N-T)\cdot M^{dc-\lfloor R_{s}/N\rfloor}+T\cdot M^{dc-\lceil R_{s}/N\rceil}$ and $R_{s}$ is used to determine the number of reference symbols, $T=(R_{s} \textbf{ mod } N)$, $\lfloor\cdot\rfloor$ and $\lceil\cdot\rceil$ are floor and ceil operators, respectively. The complexity analysis for layer nodes is similar to MPA.

For LNP SCMA, the complexity is similar to Max-log-MPA except that the constellation size $M$ is replaced by the reduced number of projections in each OFDMA subcarrier.

We summarize the decoding complexity of the above algorithms in Table~\ref{I} where $IT$ is the iteration number for MPA detection.

\section{Simulation Results}
The uplink SCMA systems with $\lambda=150\%$ and $\lambda=200\%$ are considered in this paper. SCMA codewords are designed according to~\cite{08,09} with the indicator matrix $\mathbf{F}$ defined in~\eqref{02a} and~\eqref{34} that corresponds to different overloading factors. We will compare different decoding algorithms through the BER performances and the computational complexities, respectively. The channel is Rayleigh frequency selective channel, and each tap is generated independently according to the Jakes' model with the normalized Doppler frequency $f_{d}T_{s}=0.01$, where $f_{d}$ is the maximum Doppler frequency and $T_{s}$ is the sample time. Both $4$ points and $16$ points SCMA are simulated in this paper.
\begin{equation}\label{34}
 \mathbf{F} = \left[
  \begin{array}{cccccccccccc}
  1 & 1 & 1 & 1 & 0 & 0 & 0 & 0 & 0 & 0 & 0 & 0 \\
  1 & 0 & 0 & 0 & 1 & 1 & 1 & 0 & 0 & 0 & 0 & 0 \\
  0 & 1 & 0 & 0 & 0 & 0 & 0 & 1 & 1 & 1 & 0 & 0 \\
  0 & 0 & 1 & 0 & 1 & 0 & 0 & 1 & 0 & 0 & 1 & 0 \\
  0 & 0 & 0 & 1 & 0 & 1 & 0 & 0 & 1 & 0 & 0 & 1 \\
  0 & 0 & 0 & 0 & 0 & 0 & 1 & 0 & 0 & 1 & 1 & 1 \\
  \end{array}
  \right]
\end{equation}

\begin{figure*}[th]
  %\centering
  \subfigure[Turbo SCMA 4 points, $\lambda=150\%$]{
    \label{Fig.4a} %% label for first subfigure
    \includegraphics[width=3.15in, height=2.6in]{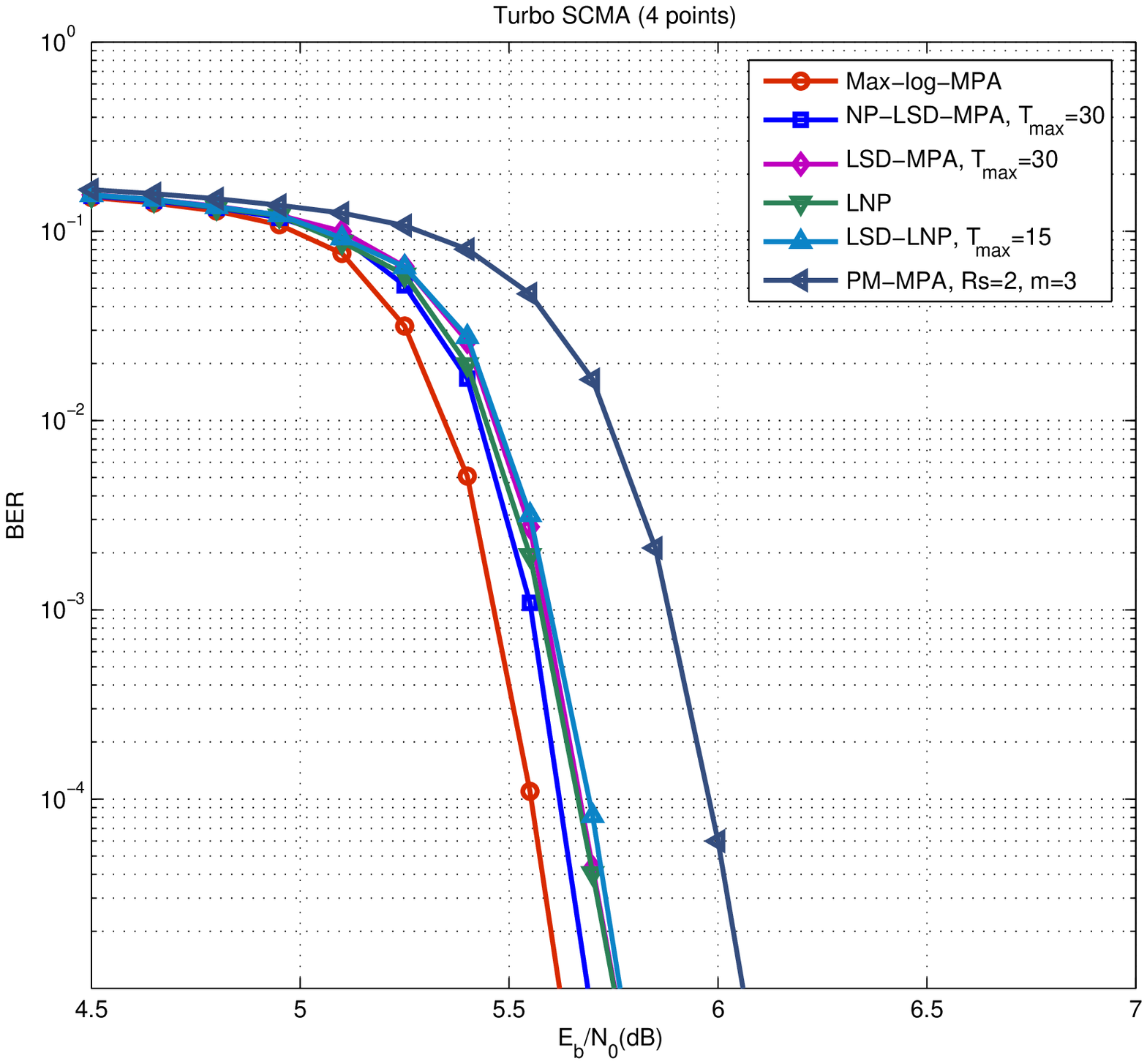}}
  \hspace{0.05in}
  \subfigure[Turbo SCMA 4 points, $\lambda=200\%$]{
    \label{Fig.4b} %% label for second subfigure
    \includegraphics[width=3.15in, height=2.6in]{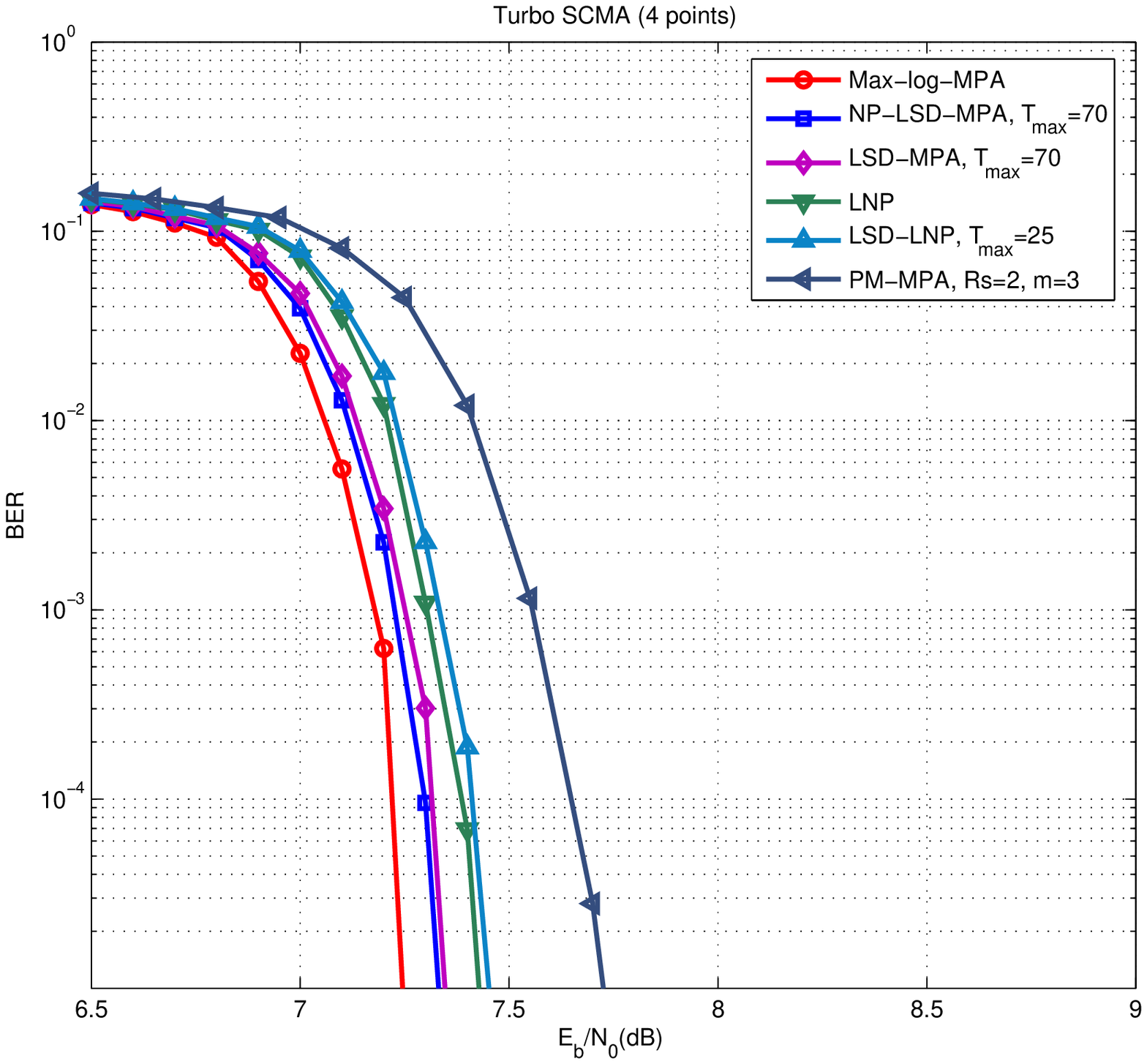}}
  \caption{BER performance of SCMA 4 points}
  \label{Fig.4} %% label for entire figure
\end{figure*}

\begin{figure*}[th]
  %\centering
  \subfigure[Turbo SCMA 16 points, $\lambda=150\%$]{
    \label{Fig.5a} %% label for first subfigure
    \includegraphics[width=3.15in, height=2.6in]{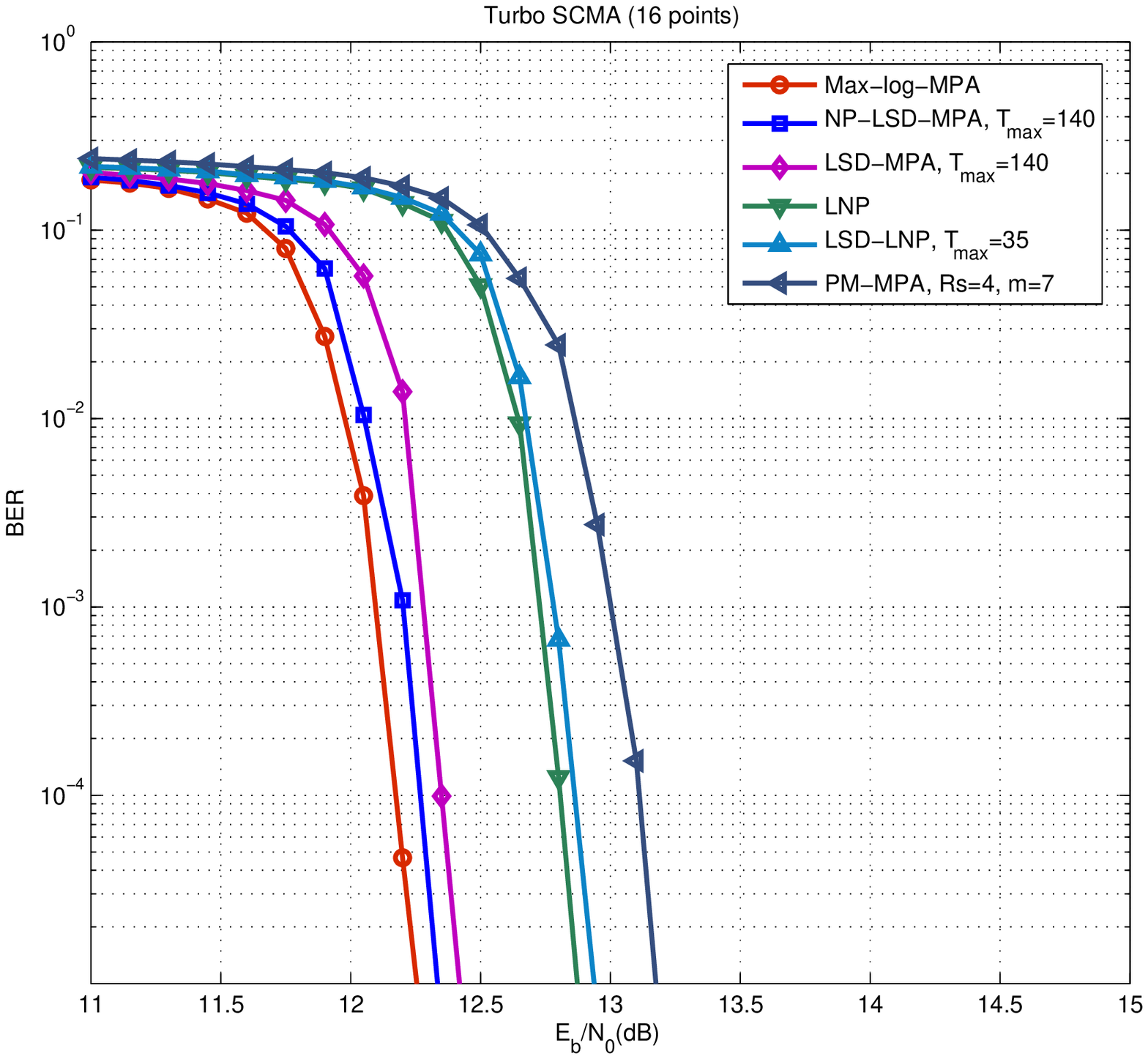}}
  \hspace{0.05in}
  \subfigure[Turbo SCMA 16 points, $\lambda=200\%$]{
    \label{Fig.5b} %% label for second subfigure
    \includegraphics[width=3.15in, height=2.6in]{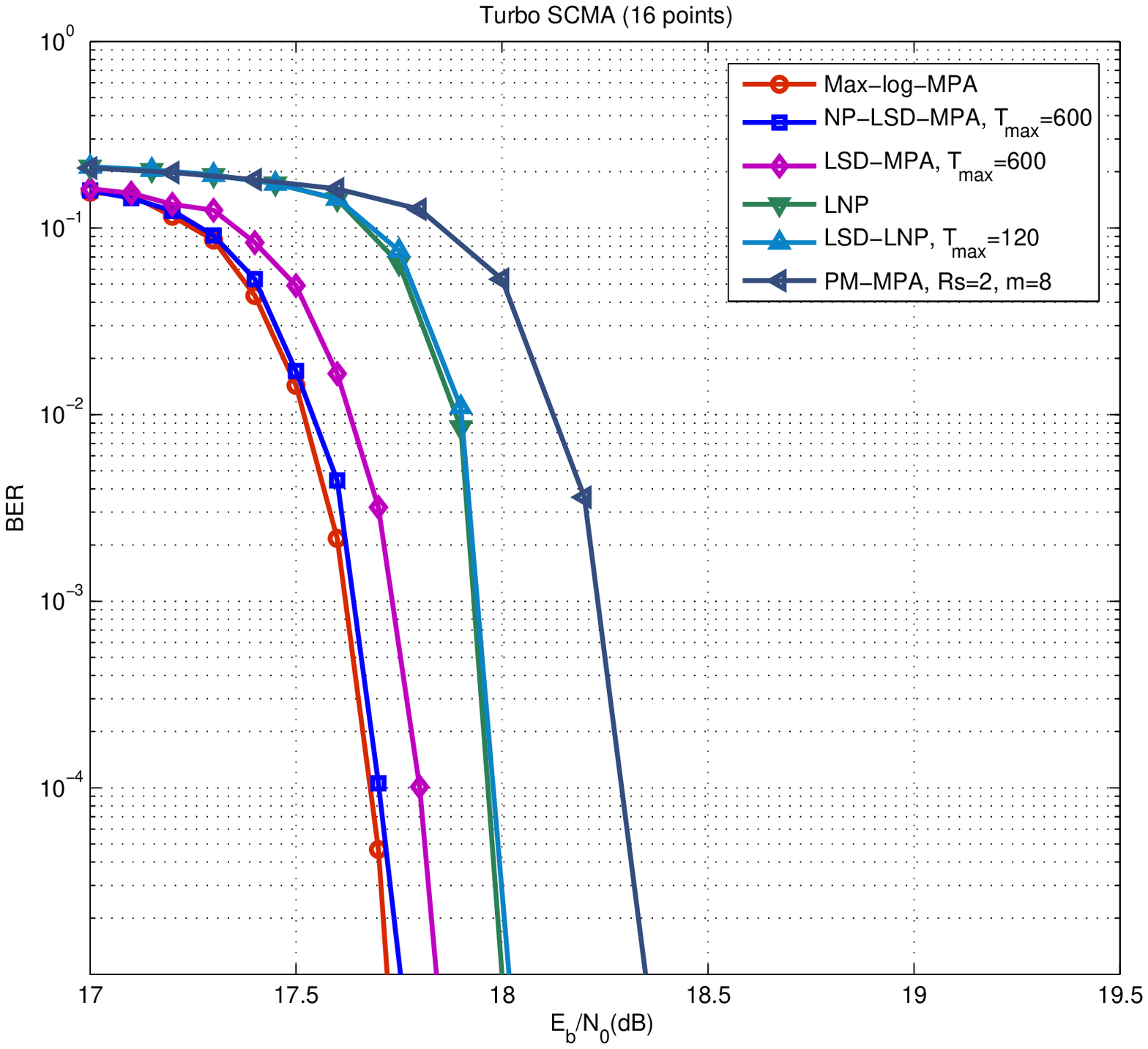}}
  \caption{BER performance of SCMA 16 points}
  \label{Fig.5} %% label for entire figure
\end{figure*}

In Fig.~\ref{Fig.4} and Fig.~\ref{Fig.5}, we show the BER performance of Max-log-MPA, LSD-MPA, NP-LSD-MPA, LNP, LSD-LNP and PM-MPA with $4$ points and $16$ points SCMA, respectively. For NP-LSD-MPA, $\epsilon$ is chosen to be $0.001$ for both $4$ points and $16$ points SCMA. The signal constellation for $4$ points LNP SCMA is demonstrated in Fig.~\ref{Fig.2a} while the signal constellation for $16$ points SCMA is demonstrated in Fig.~\ref{Fig.2b}. As in PM based MPA, $R_{s}$ is used to determine the number of referenced codewords in the last $IT-m$ iterations. Three outer loops are performed in IDD. The maximum number of MPA iteration $IT$ for $4$ points and $16$ points SCMA is set to $5$ and $10$, respectively. An $R=1/2$ rate parallel concatenated turbo code with feedforward polynomial $G_{1}=1+D+D^{3}$ and feedback polynomial $G_{2}=1+D^{2}+D^{3}$ are used as the channel coding. The interleaver size for turbo code is set to $4096$ information bits.

In LSD based MPA, we search for a candidate list set to approximate the LLR calculated in the resource node. Generally, these numbers of the list size are small enough compared with the total $M^{d_{c}}$ Euclidean distances that are required to compute in traditional MPA, e.g., for the overloading factor $\lambda=200\%$, $16$ points SCMA, since $d_{c}=4$, the total Euclidean distances needed to compute is $16^4=65536$ which is a huge number. However, by setting $T_{max}=600$ in LSD based MPA, we have $600/16^4 \approx 0.00916$, thus the searching space has been reduced effectively. Nevertheless, from the Fig.~\ref{Fig.4} and Fig.~\ref{Fig.5}, we can observe that the performance loss of NP-LSD-MPA is within $0.2$ dB for both $4$ points SCMA and $16$ points SCMA, which is negligible. Note that the NP-LSD-MPA has a better performance in BER compared with LSD-MPA. This confirms our reasoning in Section IV. The curves show in Fig.~\ref{Fig.4} and Fig.~\ref{Fig.5} confirm our prediction that the massive computation of the whole $M^{d_{c}}$ Euclidean distances is unnecessary since large Euclidean distance has tiny contribution to equation~\eqref{07}. By considering only the signal points that are within a given search radius, we can compute~\eqref{07} more efficiently and thus reduce the computational complexity of the original MPA.

In contrast, we have also simulated the BER for LNP SCMA and PM based MPA in Fig.~\ref{Fig.4} and Fig.~\ref{Fig.5}. The LNP SCMA is another way to reduce the searching space in MPA by reducing the number of projections in each subcarrier. For $\lambda=200\%$, $16$ points SCMA, if we use the constellation illustrated in Fig.~\ref{Fig.2b}, the searching space can be reduced from $16^{4}$ to $9^{4}$. The proposed LSD based MPA could also be applied to LNP SCMA and we have compared the BER performance for LSD-LNP. By the proposed method, the searching space has been further reduced from $9^{4}$ to $120$ for $\lambda=200\%$, $16$ points SCMA. Note that the constellation for LNP SCMA is generated by the suboptimal rotation matrices defined previously. Therefore, compared to the codebook that designed with the optimal rotation matrices reported in~\cite{09a}, there is some performance loss due to the suboptimal rotation of the signal constellation. The PM based MPA is also simulated. It can be observed that there is a considerable performance loss compared with Max-log-MPA especially in the high signal-to-noise ratio (SNR) region. This is due to the imprecise LLR calculated in partial marginalization since part of the codewords are justified in advance before the convergence.
\begin{figure*}
  %\centering
  \subfigure[$M=4$, $\lambda=150\%$]{
    \label{Fig.6a} %% label for first subfigure
    \includegraphics[width=3.15in, height=2.6in]{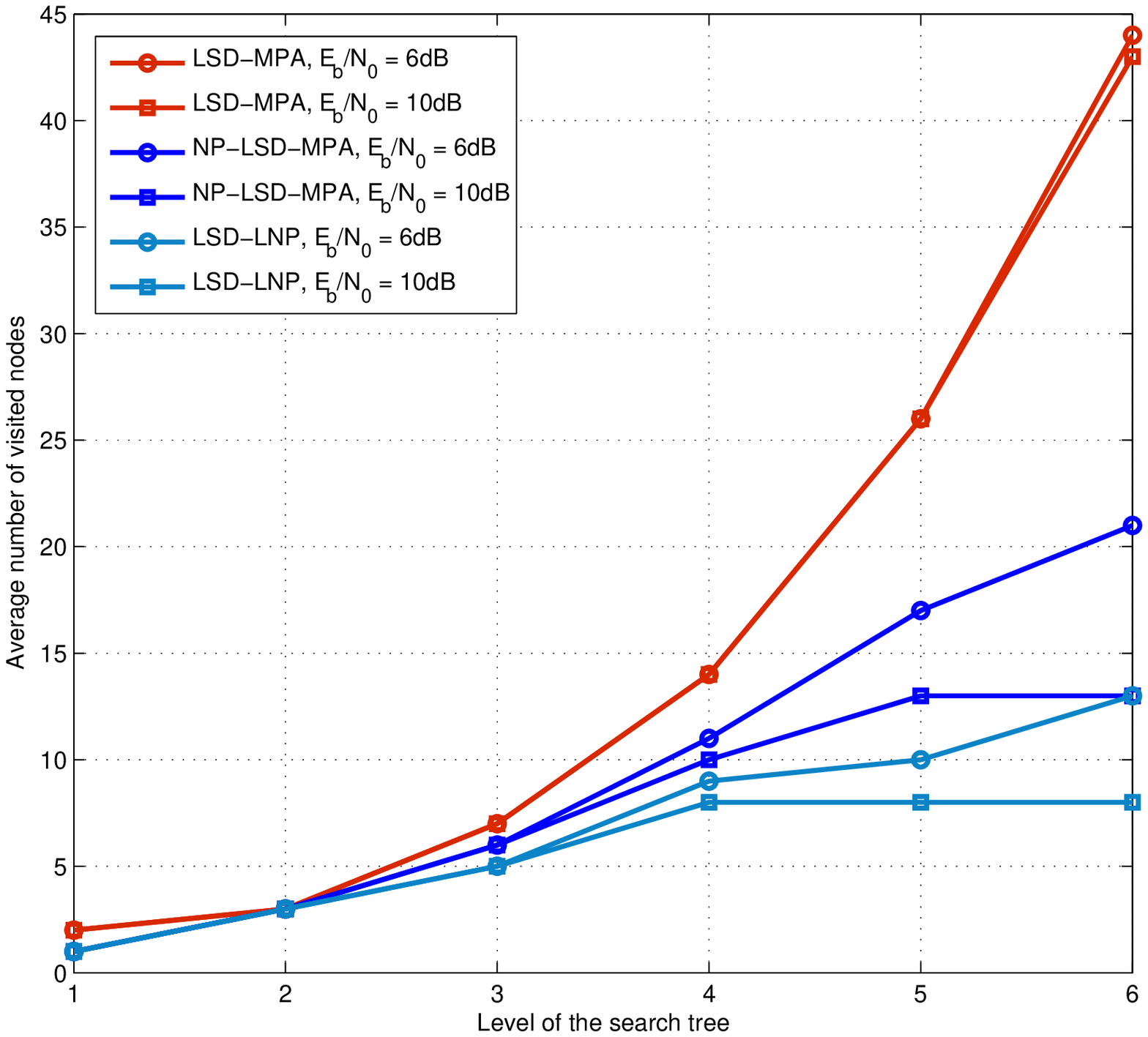}}
  \hspace{0.05in}
  \subfigure[$M=4$, $\lambda=200\%$]{
    \label{Fig.6b} %% label for second subfigure
    \includegraphics[width=3.15in, height=2.6in]{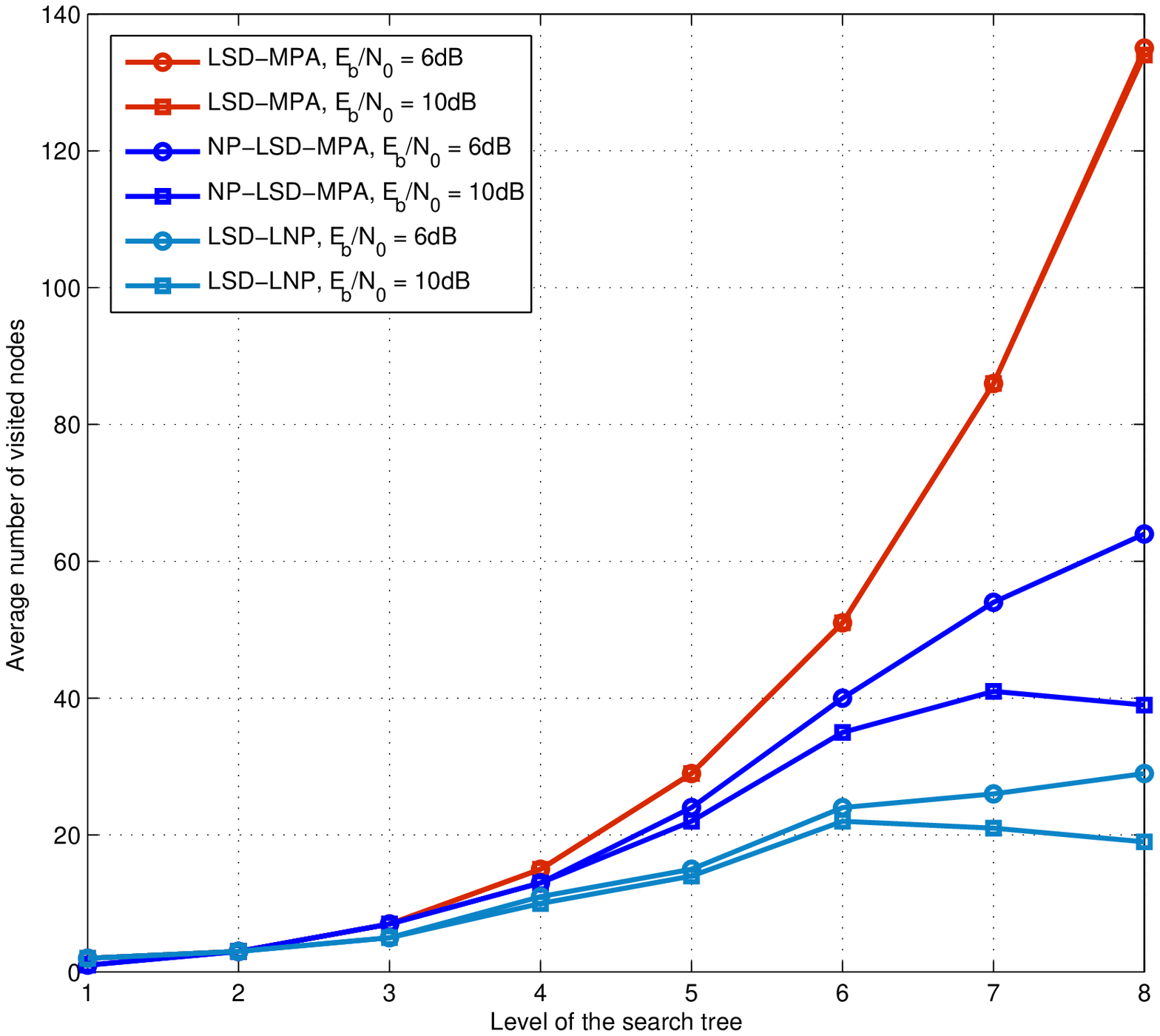}}
  \caption{Average number of nodes visited in each level (M=4)}
  \label{Fig.6} %% label for entire figure
\end{figure*}
\begin{figure*}
  %\centering
  \subfigure[$M=16$, $\lambda=150\%$]{
    \label{Fig.7a} %% label for first subfigure
    \includegraphics[width=3.15in, height=2.6in]{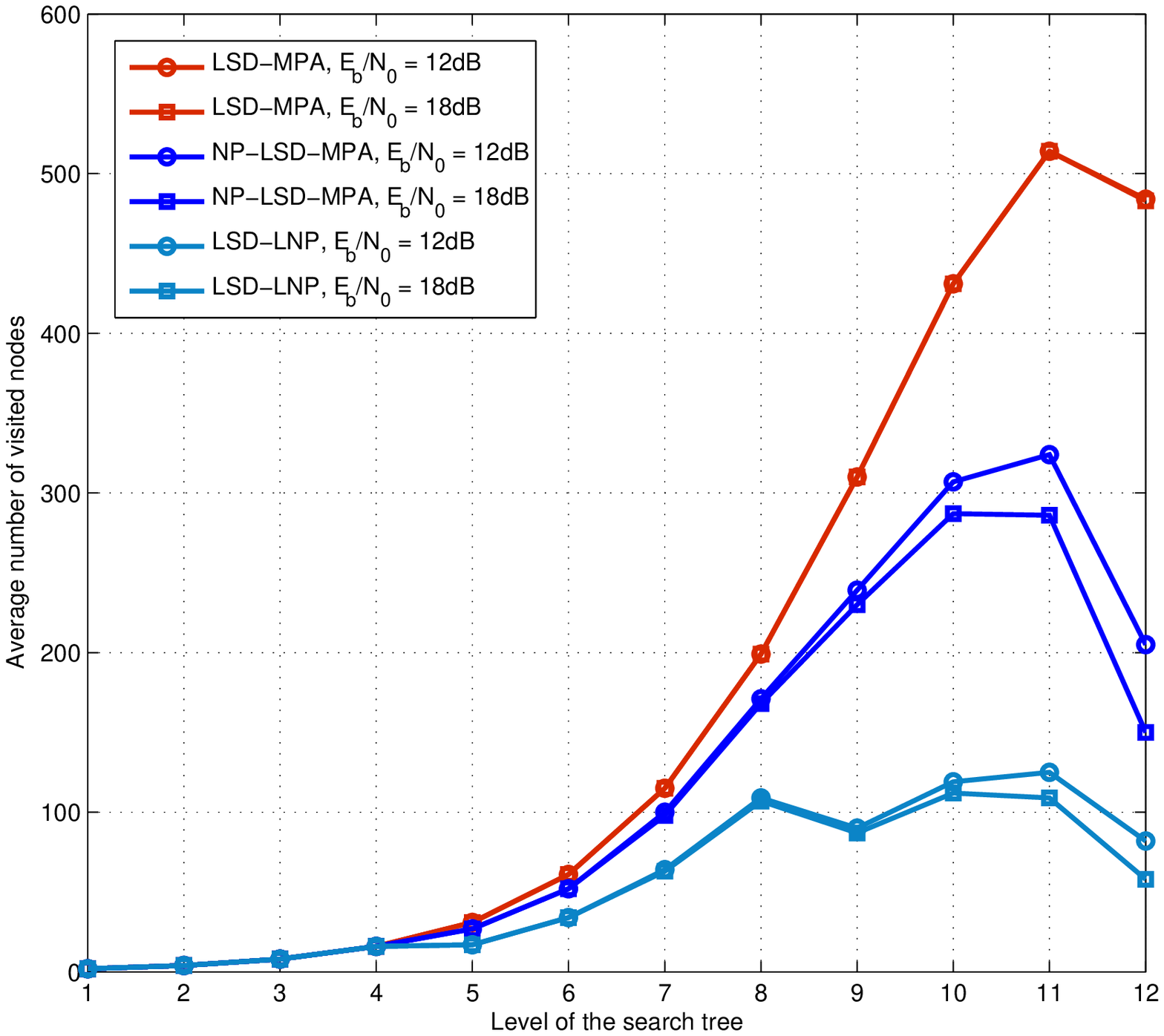}}
  \hspace{0.05in}
  \subfigure[$M=16$, $\lambda=200\%$]{
    \label{Fig.7b} %% label for second subfigure
    \includegraphics[width=3.15in, height=2.6in]{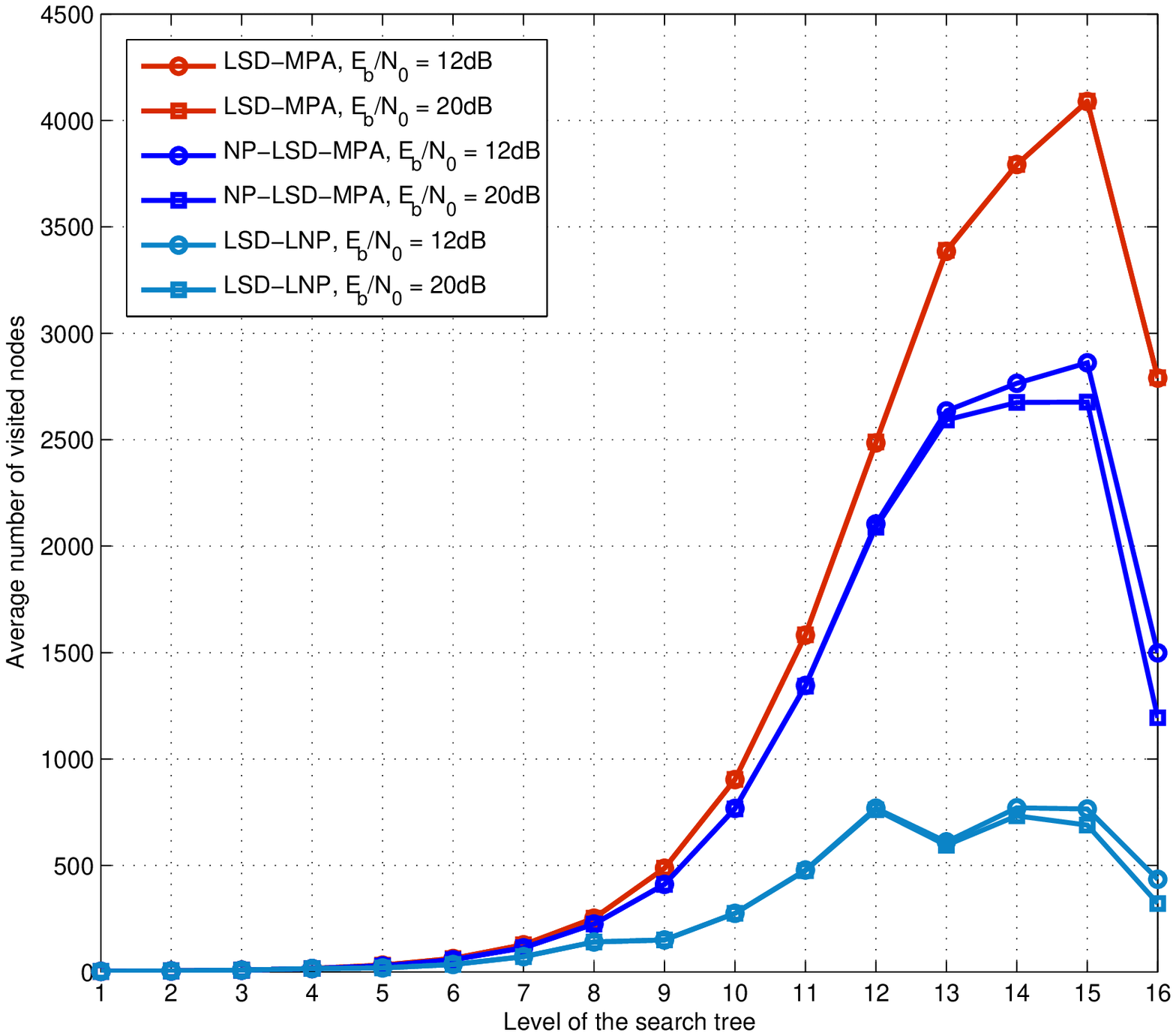}}
  \caption{Average number of nodes visited in each level (M=16)}
  \label{Fig.7} %% label for entire figure
\end{figure*}

In LSD based MPA, we reduce the decoding complexity of the conventional MPA at the expense of adding an LSD module before the MPA detector. As a tree search algorithm, the complexity of LSD depends on the averaged visited nodes during the search process. Fig.~\ref{Fig.6} and Fig.~\ref{Fig.7} illustrate the averaged visited nodes in each level of the search tree for LSD-MPA, NP-LSD-MPA and LSD-LNP. From the figures, we can observe that quite a few number of nodes are pruned due to the technique we proposed in the Section IV. In addition, the LSD-LNP has the least visited nodes among others since it has a smallest size of search tree. Note that the averaged visited nodes for LSD-MPA are invariant to SNR since the initial radius is set to infinity in SE enumeration. In contrast, with radius setup according to \eqref{32c}, the averaged visited nodes for NP-LSD-MPA and LSD-LNP decrease with the increasing of SNR. Thus the complexity will be lower in high SNR region.

\begin{figure*}
  %\centering
  \subfigure[$M=4$, $\lambda=150\%$, $E_{b}/N_{0}=5dB$]{
    \label{Fig.8a} %% label for first subfigure
    \includegraphics[width=3.15in, height=2.6in]{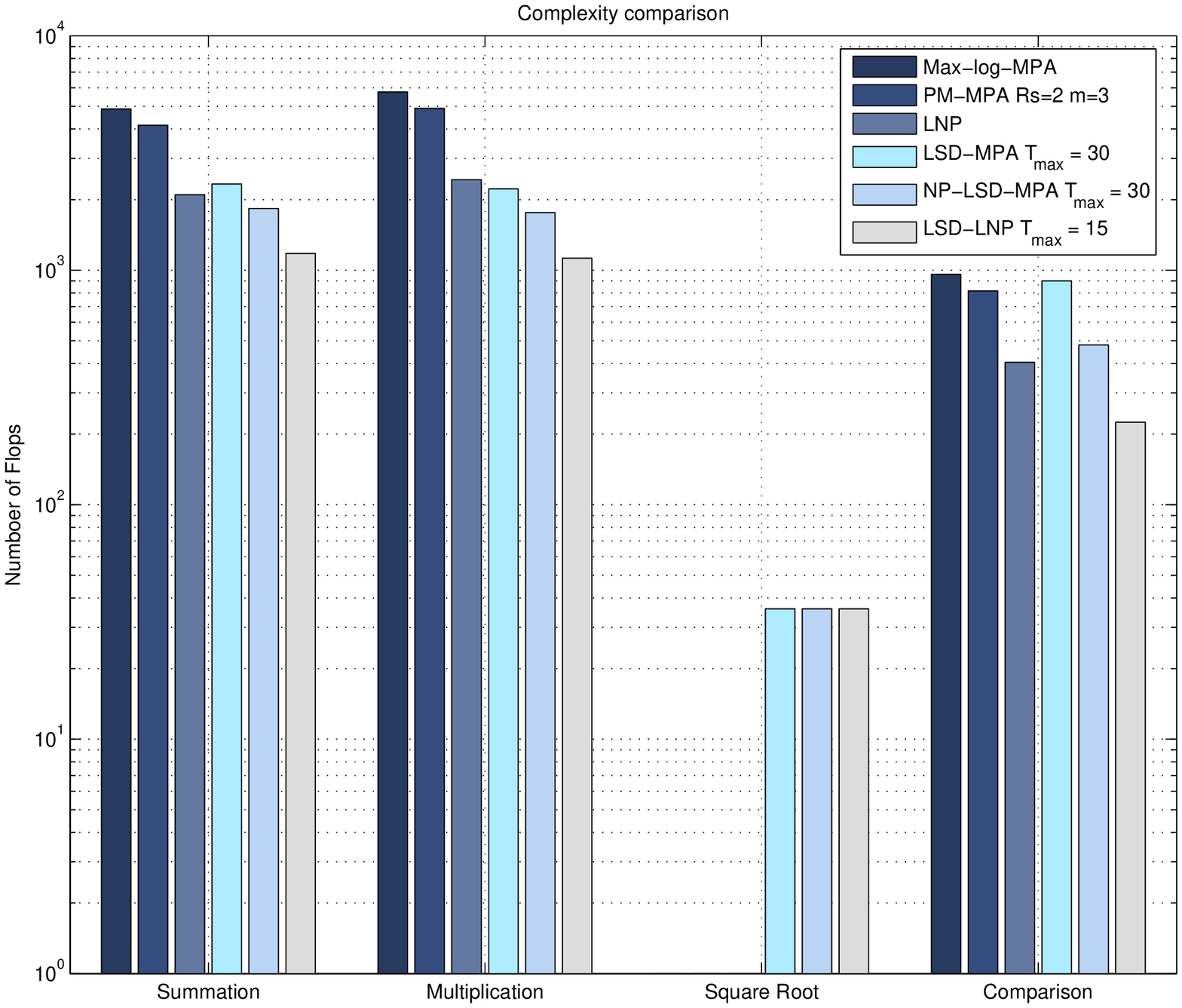}}
  \hspace{0.05in}
  \subfigure[$M=4$, $\lambda=200\%$, $E_{b}/N_{0}=7dB$]{
    \label{Fig.8b} %% label for second subfigure
    \includegraphics[width=3.15in, height=2.6in]{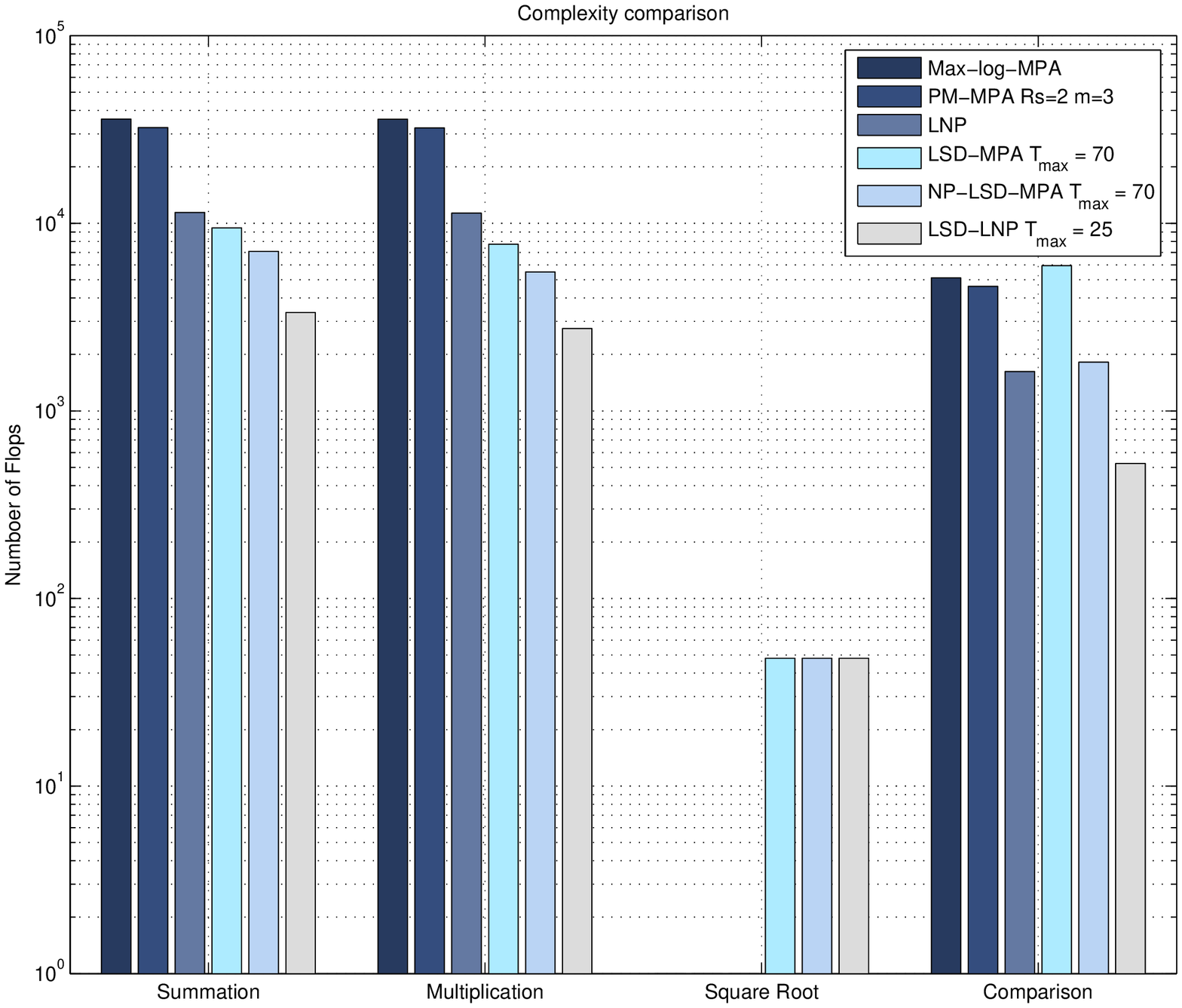}}
  \caption{The computational complexity for different decoding algorithms (M=4)}
  \label{Fig.8} %% label for entire figure
\end{figure*}

\begin{figure*}
  %\centering
  \subfigure[$M=16$, $\lambda=150\%$, $E_{b}/N_{0}=12dB$]{
    \label{Fig.9a} %% label for first subfigure
    \includegraphics[width=3.15in, height=2.6in]{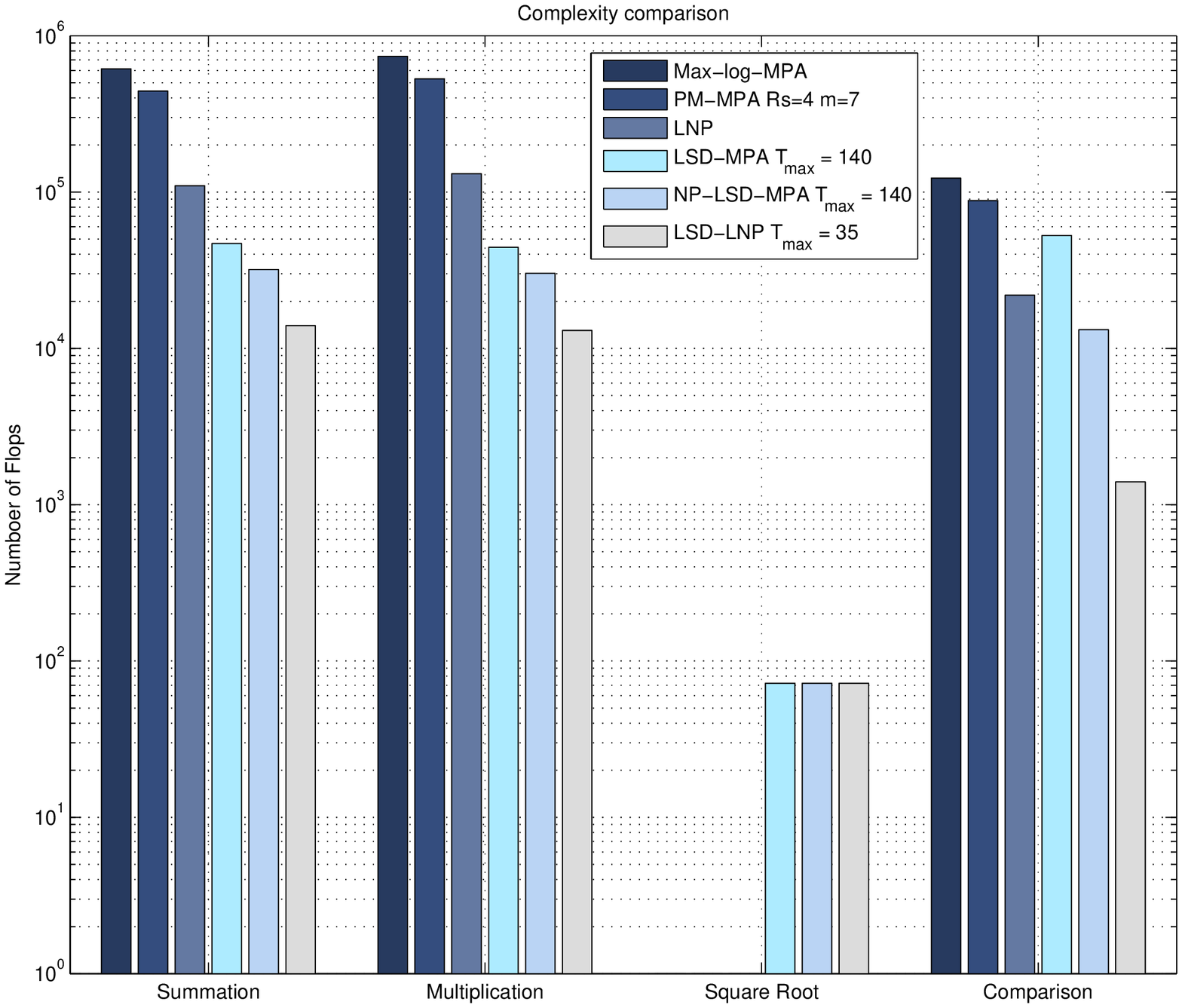}}
  \hspace{0.05in}
  \subfigure[$M=16$, $\lambda=200\%$, $E_{b}/N_{0}=17dB$]{
    \label{Fig.9b} %% label for second subfigure
    \includegraphics[width=3.15in, height=2.6in]{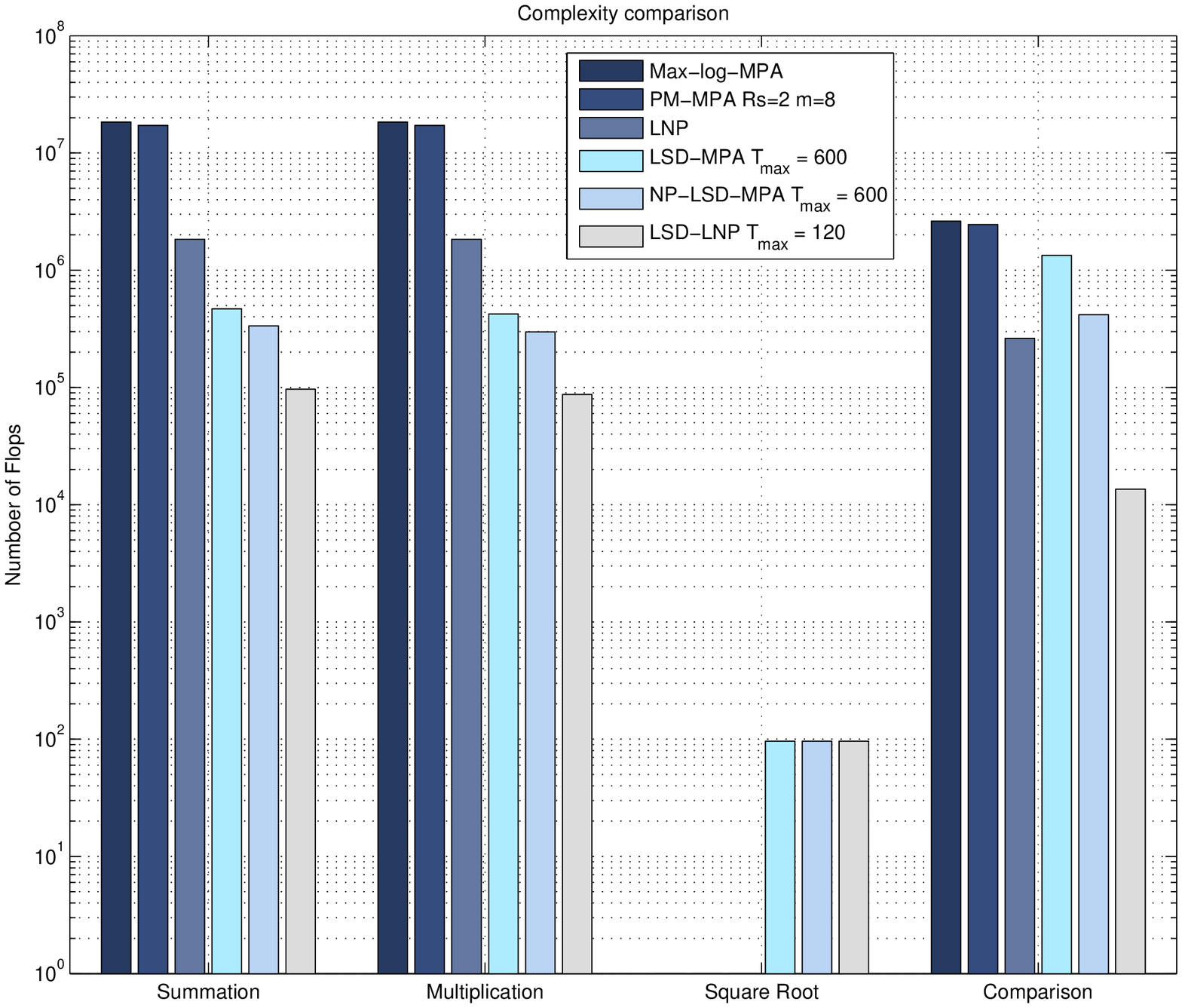}}
  \caption{The computational complexity for different decoding algorithms (M=16)}
  \label{Fig.9} %% label for entire figure
\end{figure*}

In Fig.~\ref{Fig.8} and Fig.~\ref{Fig.9}, the computational complexity is compared for different decoding algorithms. Based on the complexity analysis in Section IV and Table~\ref{I}, the histograms in the figures measure the number of arithmetic operations required by Max-log-MPA, PM-MPA, LNP, LSD-MPA, NP-LSD-MPA and LSD-LNP over one OFDMA subcarrier. From the figures, we can observe that the proposed schemes have reduced the number of summation operations and multiplication operations substantially. For $16$ points SCMA, we have reduced the FLOPs by one or more orders of magnitude. Note that some square root operations are incurred in LSD based MPA due to the implementation of the MGS QR factorization. For the square root operations, equivalent flops are measured based on IEEE floating-point representation~\cite{35}. In~\cite{35}, to compute $\sqrt{x}$, one firstly computes $y=\frac{x_{\textrm{int}}}{2}+\frac{1_{\textrm{int}}}{2}$ as a seed. Then only one Newton-Raphson iteration $\frac{1}{2}(y+\frac{x}{y})$ is used to yield a good result. The algorithm needs $6$ flops in total to complete a square root operation.

The comparison operations are also compared in Fig.~\ref{Fig.8} and Fig.~\ref{Fig.9}, it is observed that the LNP SCMA requires a less number of comparison operations compared with LSD based MPA. For LSD-MPA, the comparison operations are dominated by the search for the maximum $d(\mathbf{u}_{n})$ in step $6$ of Algorithm~\ref{alg1}. These operations can be reduced by pruning the redundancy visited nodes in the search tree. As shown in the figures, by using the node pruning techniques developed in Section IV C, the number of comparison operations for NP-LSD-MPA is reduced effectively and is comparable to LNP SCMA in general. Further, it can be observed that when combined with LSD, the LSD-LNP requires the lowest number of comparison operations among the algorithms.

\begin{table}[ht]
  \centering
  \caption{Comparison of the Computational Cost for Different Decoding Algorithms (in milliseconds)}\label{II}
  \begin{tabular}{|c||c|c|c|c|}
  \hline
   $M=16$     & Max-log-MPA & LNP & NP-LSD-MPA & LSD-LNP \\ [0.5ex] \hline\hline
  $d_{c}=3$ 　& $1.913$  & $0.2805$ & $0.2304$ & $0.1276$  \\ \hline
  $d_{c}=4$　 & $48.23$  & $7.526$  & $3.213$  & $0.9104$  \\ \hline
  \end{tabular}
\end{table}

In Table~\ref{II}, we show the per MPA iteration decoding time (in milliseconds) for $\lambda=150\%$ and $\lambda=200\%$, $16$ points SCMA. The SNR is set to $12dB$ for $\lambda=150\%$ SCMA while $17dB$ for $\lambda=200\%$ SCMA. It can be observed that using the proposed NP-LSD-MPA, the computational cost has been reduced more than $90\%$ compared with the original MPA detection which is a notable reduction on the computational cost. Further, the combination of LSD and LNP could attain an even more remarkable result. Therefore, the proposed algorithm is effective on reducing the SCMA decoding complexity.

In the last part of the simulation, the impact of channel uncertainty on the BER performance is investigated. The multipath fading channel state information (CSI) is modeled as,
\begin{equation}\label{35}
  h = \widetilde{h} + \Delta h,
\end{equation}
where $h$ and $\widetilde{h}$ are the actual CSI and the estimated CSI, respectively, while $\Delta h$ is the corresponding CSI error which is modeled as a Gaussian noise and is independent of actual CSI. Here, $h$, $\widetilde{h}$ and $\Delta h$ are assumed to have zero means and normalized variances of $1$,$1-\xi$ and $\xi$, respectively.

\begin{figure*}
  %\centering
  \subfigure[$\lambda=150\%$]{
    \label{Fig.10a} %% label for first subfigure
    \includegraphics[width=3.15in, height=2.6in]{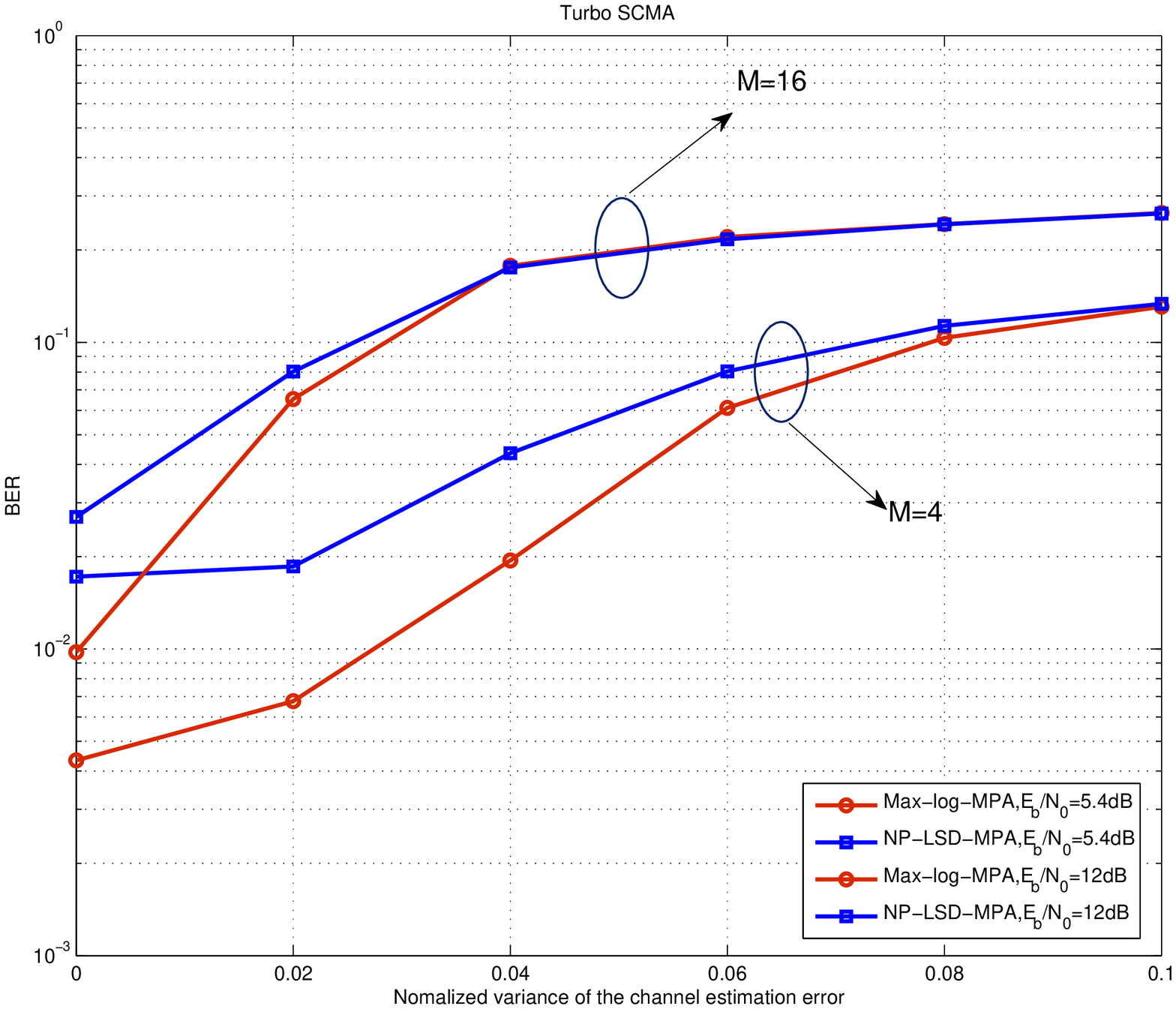}}
  \hspace{0.05in}
  \subfigure[$\lambda=200\%$]{
    \label{Fig.10b} %% label for second subfigure
    \includegraphics[width=3.15in, height=2.6in]{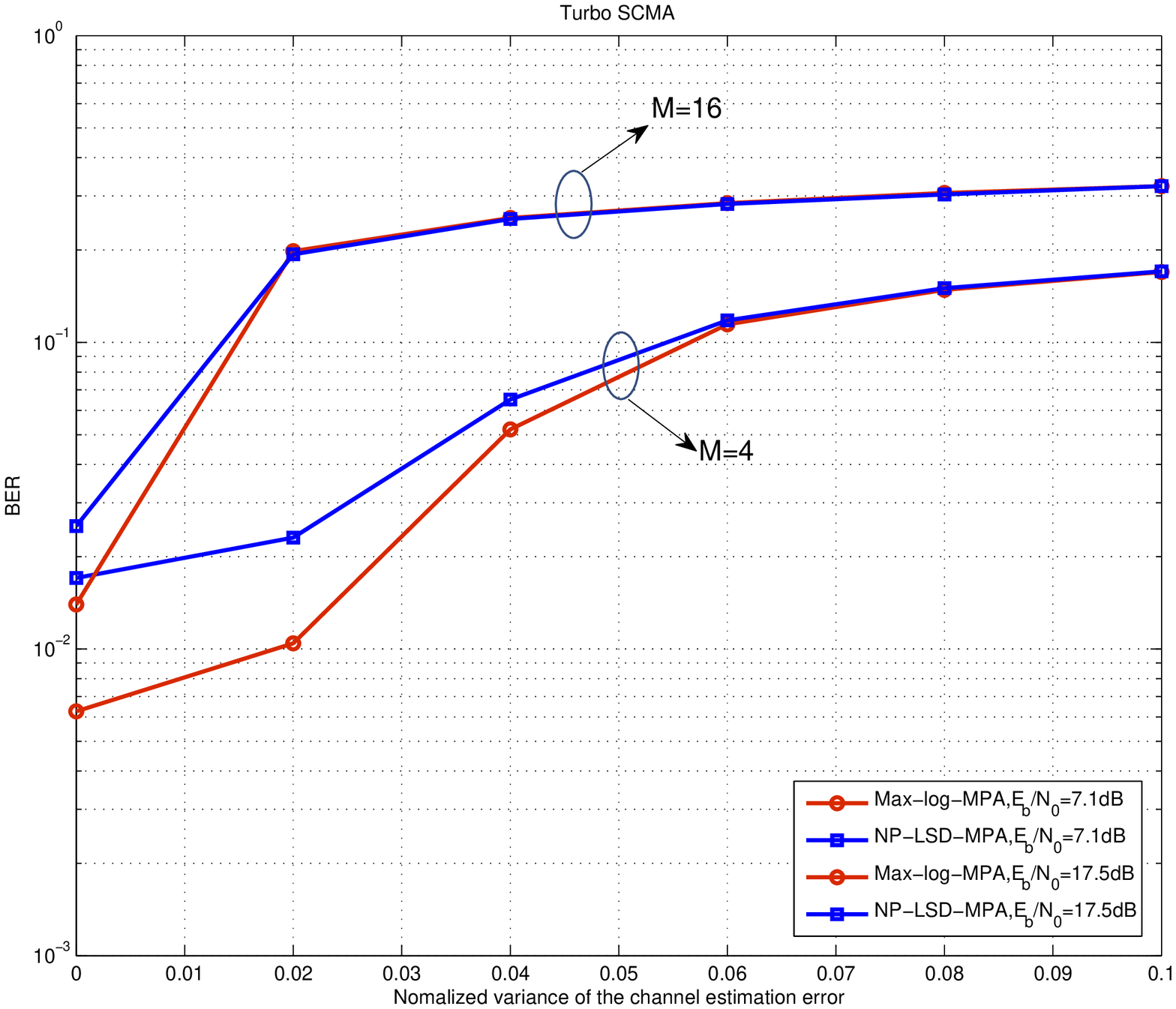}}
  \caption{The effects of the channel estimation error on the BER performance}
  \label{Fig.10} %% label for entire figure
\end{figure*}

From Fig.~\ref{Fig.10}, we can observe that when the channel estimation error $\xi$ is small, there is a gap between the Max-log-MPA and NP-LSD-MPA. This is because the LLRs calculated in MPA is the approximation by using the candidate set in LSD based MPA. In this case, the errors in the LLR approximation is the main factor to the BER performance and one can increase the SNR (within $0.2$dB typically as observed in Fig.~\ref{Fig.4} and Fig.~\ref{Fig.5}) to remove this gap. As the channel estimation error increases, however, the gap shrinks and both the algorithms have a deteriorated performance due to the imperfect channel state information at the receiver side. In this case, channel estimation error is the main factor to the BER performance. A robust design for SCMA decoder will be the future research direction.

\section{Conclusion}
In this paper, a low complexity SCMA decoder based on list sphere decoding is introduced to reduce the computational complexity for MPA detection. We first discuss the encoding of SCMA codewords and show that they are essentially complex lattice constellation points. By exploring the lattice structure of the codewords, the LSD is implemented before the MPA detection. In general, the LSD finds all possible hypotheses within a hypersphere, thus avoids the exhaustive search in ML detection. As the complexity of LSD depends on the averaged visited nodes during the search process, some node pruning techniques are further developed to reduce the complexity of LSD. Simulation results show that the proposed algorithm can attain a near ML performance with a substantially reduced decoding complexity compared with original MPA detection or other decoding methods.

\ifCLASSOPTIONcaptionsoff
  \newpage
\fi

\end{document}